\documentclass[preprint]{elsarticle}
\usepackage{hyperref}
\usepackage{amsmath}
\usepackage{amssymb}
\usepackage{color}
\usepackage{graphicx}
\usepackage{algorithm}
\usepackage{algorithmicx}
\usepackage{algpseudocode}
\usepackage{listings}
\usepackage{multicol}
\usepackage{multirow}
\usepackage{lineno}
\usepackage{setspace}

\journal{Computer Physics Communications}

\newcommand{\figref}[1]{Figure~\ref{fig:#1}}
\newcommand{\Rey}{\mathrm{Re}}
\newcommand{\ket}[1]{\left| #1 \right>} 
\newcommand{\lstref}[1]{Listing~\ref{lst:#1}}
\renewcommand{\algref}[1]{Algorithm~\ref{alg:#1}}
\newcommand{\feq}{f^\mathrm{eq}}

\begin{document}

\begin{frontmatter}
\title{Sailfish: a flexible multi-GPU implementation of the lattice Boltzmann method}
\author[us,google]{M. Januszewski}
\ead{michalj@gmail.com}

\author[us]{M. Kostur\corref{cor1}}
\ead{marcin.kostur@us.edu.pl}
\address[us]{Institute of Physics, University of Silesia, 40-007 Katowice, Poland}
\address[google]{Google Switzerland GmbH, Brandschenkestrasse 110, 8002 Zurich, Switzerland}
\cortext[cor1]{Corresponding author}

\begin{abstract}
We present Sailfish, an open source fluid simulation package implementing
the lattice Boltzmann method (LBM) on modern Graphics Processing Units (GPUs) using CUDA/OpenCL.  We take a novel
approach to GPU code implementation and use run-time code generation techniques and
a high level programming language (Python) to achieve state of the
art performance, while allowing easy experimentation with different LBM models and tuning for
various types of hardware.  We discuss the general design principles of the code, scaling
to multiple GPUs in a distributed environment, as well as the GPU implementation and optimization of many
different LBM models, both single component (BGK, MRT, ELBM) and multicomponent (Shan-Chen, free energy).
The paper also presents results of performance benchmarks spanning the
last three NVIDIA GPU generations (Tesla, Fermi, Kepler), which we hope will be useful
for researchers working with this type of hardware and similar codes.
\end{abstract}

\begin{keyword}
lattice Boltzmann
\sep LBM
\sep computational fluid dynamics
\sep graphics processing unit
\sep GPU
\sep CUDA
\end{keyword}

\end{frontmatter}


{\bf PROGRAM SUMMARY}

\begin{small}
\noindent
{\em Manuscript Title: } Sailfish: a flexible multi-GPU implementation of the lattice Boltzmann method                                     \\
{\em Authors:} Michal Januszewski, Marcin Kostur                                                \\
{\em Program Title:} Sailfish                                       \\
{\em Code Repository:} \url{https://github.com/sailfish-team/sailfish} \\
{\em Journal Reference:}                                      \\
{\em Catalogue identifier:}                                   \\
{\em Licensing provisions: LGPLv3}                                   \\
{\em Programming language:} Python, CUDA C, OpenCL                                   \\
{\em Computer:} any with an OpenCL or CUDA-compliant GPU                                               \\
{\em Operating system:} no limits (tested on Linux and Mac OS X)                                       \\
{\em RAM:} Hundreds of megabytes to tens of gigabytes for typical cases.                                              \\
{\em Keywords:} lattice Boltzmann, LBM, CUDA, OpenCL, GPU, computational fluid dynamics, Python.  \\
{\em Classification:} 12, 6.5                                        \\
{\em External routines/libraries:} PyCUDA/PyOpenCL, Numpy, Mako, ZeroMQ (for multi-GPU simulations)                       \\
{\em Nature of problem:}\\
GPU-accelerated simulation of single- and multi-component fluid flows.
   \\
{\em Solution method:}\\
A wide range of relaxation models (LBGK, MRT, regularized LB, ELBM, Shan-Chen,
free energy, free surface) and boundary conditions within the lattice Boltzmann method framework.
Simulations can be run in single or double precision using one or more GPUs.
   \\
{\em Restrictions:} \\
The lattice Boltzmann method works for low Mach number flows only.
\\
{\em Unusual features:}\\
The actual numerical calculations run exclusively on GPUs. The numerical code
is built dynamically at run-time in CUDA C or OpenCL, using templates and symbolic formulas.
The high-level control of the simulation is maintained by a Python process.
   \\
{\em Running time:} problem-dependent, typically minutes (for small cases or short simulations)
to hours (large cases or long simulations)
   \\
\end{small}

\section{Introduction}

Merely a few years ago the computational power on the order of TFLOPS was an
attribute of a supercomputer.  Today, the latest commodity Graphics Processing
Units (GPU) are capable of more than 5 TFLOPS  
and have much lower energy requirements than a set of Central Processing Units (CPUs)
of comparable performance -- a result achieved by employing a massively parallel architecture
with hundreds of cores on a single device.

In order to take advantage of the parallel hardware, appropriate algorithms
have to be developed, and this generally needs to be done on a problem-specific
basis.  Ideally, the problem naturally decomposes into a large number of (semi-)independent tasks
whose results can be combined in a simple way, as in e.g. numerical Monte-Carlo solution of
a stochastic differential equation~\cite{Januszewski2010}, parameter space studies of dynamical
systems, or event simulation and reconstruction in particle physics~\cite{halyo2013gpu}.

In recent years, the lattice Boltzmann method (LBM) emerged as an interesting
alternative to more established methods for fluid flow simulations.  Originally developed as an extension
of lattice gas automata, nowadays LBM stands on well-established theoretical
foundations and serves as a method of choice for many researchers
in various fields~\cite{ChenDoolen1998, Aidun2010rev}, while still retaining its relative simplicity.
Perhaps its most important feature in practice is its suitability for parallel
architectures -- the algorithm is executed on a regular lattice and
typically only interactions between nearest neighbor nodes are necessary.

This paper discusses various issues related to the implementation of the LBM on GPUs,
and presents a concrete and flexible solution, taking a new overall approach
to the problem of LB software design. Our \textit{Sailfish} code has been under development since April 2009 and is
publicly available under an open source license.  With support for both single
and binary fluid simulations, a wide variety of boundary conditions, and calculations
on both a single GPU and multiple GPUs, it is, to the best of our knowledge, the most comprehensive
open source LBM code for GPUs.

The text is organized as follows.  First,
a short overview of the discussed lattice Boltzmann models is presented.  Then,
the primary design principles of the Sailfish project are laid out, and various
LB GPU implementations are compared.  This is followed by performance measurements
on diverse GPU hardware, both on single machines and in clusters.  In the next
section, the code is validated on 4 standard test cases.  The final section
concludes the paper and provides some directions for future work.

\section{Lattice Boltzmann methods overview}

In this section, \emph{all} models implemented in Sailfish as of September 2013
will be briefly overviewed.
We start with the basic concepts and nomenclature of LBM. We then describe single
fluid models: single-relaxation time LBGK and regularized dynamics, multiple relaxation
times and the entropic LBM, and multi-component models: the Shan-Chen model and the
free energy model. Finally we give a short overview of our implementation of body forces
and various boundary conditions. We refer the reader to the reviews
\cite{ChenDoolen1998, Aidun2010rev} for a more detailed discussion of the LBM.

\begin{figure}
	\centering
	\includegraphics[width=\textwidth]{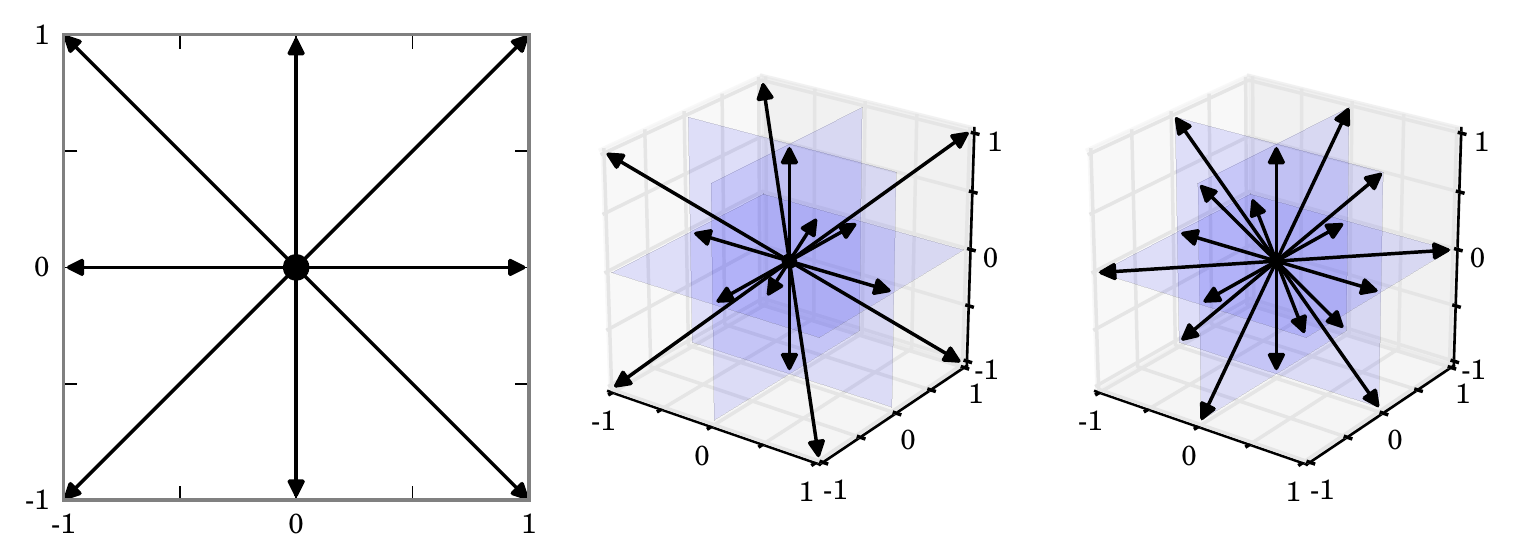}
	\caption{Sample lattices used for LB simulations: D2Q9 (left panel), D3Q15 (middle panel), and D3Q19 (right panel).
Sailfish additionally implements D3Q27 and D3Q13~\cite{PhysRevE.63.066702}.}
	\label{fig:lattices}
\end{figure}

In an LB simulation space is discretized into a regular Cartesian grid, where each node
represents a small pocket of fluid. The state of the fluid is encoded by a particle
distribution function $f_i$ where $i$ spans a set of discrete velocity vectors $\{ \vec{e}_i \}$
indicating the allowed directions of mass movement between the nodes of the lattice.
$f_i$ are often also called \textit{mass fractions}. LB lattices are typically named
using the DxQy scheme, where $x$ indicates the dimensionality
of the lattice, and $y$ is the number of discrete velocity vectors. \figref{lattices} shows
some common LB lattices.

Macroscopic fluid fields, such as density ($\rho$) or velocity ($\vec{u}$) are defined as moments
of the distribution function:
\begin{equation}
	\rho = \sum_i f_i, \qquad \rho \vec{u} = \sum_i f_i \vec{e}_i
	\label{eq:lb_macro}
\end{equation}

In the simplest case the system dynamics is described by:
\begin{equation}
	f_i(\vec{x} + \vec{e}_i, t + 1) - f_i(\vec{x}, t) = \frac{1}{\tau}\left(f_i - f_i^{\mathrm{eq}} \right)(\vec{x}, t)
	\label{eq:lbgk}
\end{equation}
where $\left\{ f_i^{\mathrm{eq}}(\vec{x}, t), i = 1, \ldots,  y \right\}$ is a set of equilibrium distributions which are functions of the
macroscopic fields ($\rho$, $\vec{u}$) at node $\vec{x}$ at time $t$, $\tau$ is a relaxation
time related to the kinematic viscosity $\nu$ via $\tau = (1 + 6 \nu) / 2$, and where the positions
and time are expressed in the lattice unit system, i.e. the time unit represents a single
simulation step and the lattice nodes are separated by 1 distance unit along the primary
axes. This LB model is often referred to as
Lattice Bhatnagar-Gross-Krook (LBGK), which owes its name to the BGK operator
from kinetic theory \cite{bgk54}, and by which the collision operator on the
right hand side of \eqref{eq:lbgk} is inspired.

\begin{table}
	\caption{Weights $w_i$ for different lattices implemented in Sailfish.}
	\centering
	\begin{tabular}{l | c c c c}
		\hline
		\multirow{2}{*}{Lattice} & \multicolumn{4}{c}{Link length} \\
		& 0 & 1 & $\sqrt{2}$ & $\sqrt{3}$ \\
		\hline
		D2Q9 & 4/9 & 1/9 & 1/36 & - \\
		D3Q13 & 1/2 & - & 1/24 & - \\
		D3Q15 & 2/9 & 1/9 & - & 1/72 \\
		D3Q19 & 1/3 & 1/18 & 1/36 & -  \\
		D3Q27 & 8/27 & 2/27 & 1/54 & 1/216 \\
		\hline
	\end{tabular}
	\label{tbl:bgkweights}
\end{table}

When $f_i^{\mathrm{eq}}$ takes the form:
\begin{equation}
	f_i^{\mathrm{eq}} = \rho w_i \left( 1 + 3 \vec{e}_i \cdot \vec{u} + \frac{9}{2} (\vec{e}_i \cdot \vec{u} )^2 -
		\frac{3}{2} u^2 \right)
	\label{eq:equilibrium}
\end{equation}
where $\{w_i\}$ is a set of weights (see Table~\ref{tbl:bgkweights}), the Navier-Stokes equations can be recovered from \eqref{eq:lbgk}
in the limit of low Mach numbers through Chapman-Enskog expansion \cite{ChenDoolen1998} or expansion in a Hermite
basis \cite{ShanHe1998}. It can be shown that \eqref{eq:lbgk}
is second-order accurate in space and time in the bulk of the fluid~\cite{ChenDoolen1998}.

It should be noted that \eqref{eq:lb_macro} and the right hand side of \eqref{eq:lbgk} are
fully local, i.e. only information from node $\vec{x}$ is necessary to calculate the system
state at the next step of the simulation.  When data is exchanged between nodes, the process
is limited to the nearest neighbors (left hand side of \eqref{eq:lbgk}).  This
makes the LB method ideally suited for implementation on massively parallel computer architectures.

Conceptually, \eqref{eq:lbgk} is often decomposed into two steps: relaxation (evaluation of
the collision operator, right hand side) and propagation (movement of mass fractions to neighboring nodes, left hand side).  While
such a decomposition is often suboptimal from the point of view of a software implementation,
it still provides a useful mental framework to discuss LB algorithms, and we will use these
two names throughout the text.

Many extended LB models have been proposed to (a) to improve numerical stability, or (b) simulate
more complex systems. A comprehensive discussion of such models would be prohibitively long,
so here we limit ourselves to a short overview of models that are currently implemented within the Sailfish
framework. In the first group, we discuss the multiple-relaxation times (MRT) model,
the entropic LB model (ELBM) and the Smagorinsky large eddy model. In the second group, Sailfish implements two multicomponent
fluid models: the Shan-Chen model \cite{ShanChen1993}, and the free energy model \cite{swift-binary}.

\subsection{Single fluid models with enhanced stability}

One practical problem with the LBGK approximation is the limited range of viscosities at which
the model remains stable, which directly translates into a limited range of Reynolds numbers
$\Rey = L u / \nu$ at which fluid flows can be simulated ($L$ being the spatial extent of the domain
expressed as a number of nodes). With $u$ limited to about $0.05$ by the low Mach number requirement
for most LB models where the speed of sound $c_s = 1/\sqrt{3}$ and the stable range of numerical viscosities
at these speeds being about $(10^{-3}, 1/6)$, the highest attainable Reynolds number assuming $L = 300$ is on the order
of $\Rey = 15000$. In practice, the range of Reynolds numbers and viscosities is often even narrower due to
instabilities introduced by complex geometry or boundary conditions. To address this problem, a number
of models exhibiting enhanced numerical stability have been developed.

\subsubsection{Regularized LBM}

Perhaps the simplest modification of the LBGK model aiming at improving its stability is the
regularized model~\cite{latt_regularized}. The basic idea of this approach is to replace the nonequilibrium
part of the distributions $f_i^{\mathrm{neq}} = f_i - \feq_i$ with a new first order regularized value $f^{\mathrm{reg}}$
derived from the non-equilibrium stress tensor $\Pi_{\alpha \beta}^{\mathrm{neq}} = \sum_i f_i^{\mathrm{neq}} (e_{i\alpha} e_{i \beta} - \delta_{\alpha \beta} c_s^2)$ as 
$f^{\mathrm{reg}} = \frac{w_i}{2 c_s^4} (e_{i\alpha} e_{i\beta} - c_s^2 \delta_{\alpha \beta}) \Pi_{\alpha \beta}^{\mathrm{neq}}$.
After this regularization step, the collision and propagation proceed as in the standard LBGK scheme.

\subsubsection{Multiple relaxation times}

If the particle distributions $f_i$ are considered to form vectors $\ket{f}$, the collision
operator in the LBGK approximation \eqref{eq:lbgk} can be interpreted as a diagonal
matrix $S = \omega I$, with $\omega = 1/\tau$ acting on the vector difference
$\ket{\feq(\vec{x}, t)} - \ket{f(\vec{x}, t)}$. The basis of the distribution vector can
be changed using an invertible matrix $M$: $\ket{m} = M \ket{f}$. In particular, the matrix $M$
can be chosen such that $\ket{m}$ is a vector of moments of the particle distributions $f_i$.

The idea of MRT is to perform the collision in the moment basis, and then change the basis
to recover the original form of the particle distributions for the streaming step \cite{DhumieresMrt2002}.
The MRT collision operator can be written as:
\begin{equation}
	M^{-1} \hat{S} \left( \ket{m^{\mathrm{eq}}(\vec{x}, t)} - \ket{m(\vec{x}, t)}\right)
	\label{eq:mrt_collision}
\end{equation}
where the collision matrix $\hat{S} = M S M^{-1}$ is diagonal with its elements $s_i = 1 / \tau_i$
representing the inverses of relaxation times for individual moments. Collision matrix elements
corresponding to conserved moments, such as density or momentum are set to 0. Some of the remaining
elements are related to kinematic and bulk viscosities, but the other ones can be tuned
to increase the numerical stability of the simulation without changing its physical results.
We refer the reader to the paper of D'Humieres et al.~\cite{DhumieresMrt2002}
for a detailed discussion of optimal choices for relaxation times, and the form of the matrix $M$.

\subsubsection{Entropic LBM}

In the entropic LB model, a discrete H-function $H(f) = \sum_i f_i \ln (f_i / w_i)$ is
defined and the equilibrium distribution function is defined as the extremum of $H(f)$
under constraints of mass and momentum conservation:
\begin{equation}
	f_i^{\mathrm{eq}} = \rho w_i \prod_{\alpha=1}^D \left( 2 - \sqrt{1 + u_{\alpha}^2} \right) \left( \frac{2 u_{\alpha} + \sqrt{1 + 3u_{\alpha}^2}}{1 - u_{\alpha}} \right)^{e_{i \alpha}}
\end{equation}
where $D$ is the dimensionality of the lattice.
The relaxation process is also modified to include a new dynamically adjusted parameter $\alpha$:
\begin{equation}
	f_i(\vec{x} + \vec{e}_i, t + 1) - f_i(\vec{x}, t) = \omega_0 \frac{\alpha}{2} \left(f_i - f_i^{\mathrm{eq}} \right)(\vec{x}, t)
	\label{eq:elbm}
\end{equation}
with $\omega_0 = 1 / \tau$. $\alpha$ is evaluated at every simulation step as the solution of
the $H$-function monotonicity constraint:
\begin{equation}
	H(f) = H\left(f - \alpha (f - \feq)\right)
	\label{eq:alpha_cond}
\end{equation}
This procedure guarantees unconditional stability of the numerical scheme, as the
$H(f)$ function can be proven to be a Lyapunov function for the system.
When $\alpha = 2$ as is the case when the system is close
to equilibrium, \eqref{eq:elbm} has the same form as \eqref{eq:lbgk}. The corrections
resulting from H-theorem compliance can be both positive and negative, temporarily
increasing and decreasing the effective local viscosity of the fluid~\cite{ansumali_minimal_2007,chikatamarla_entropic_2006}.

\subsubsection{Smagorinsky subgrid model}

The idea behind the Smagorinsky subgrid model is to locally modify the fluid viscosity
by adding an eddy viscosity term dependent on the magnitude of the strain rate tensor $S$:
\begin{equation}
	\nu = \nu_0 + \nu_t
	\label{eq:smagorinsky}
\end{equation}
where $\nu_0$ is the baseline viscosity of the fluid and $\nu_t = \tau_t / 3$ is an eddy viscosity,
the form of which depends on the subgrid model used in the simulation. In the Smagorinsky
model, the relaxation time $\tau_t$ can be calculated using the momentum flux tensor
$Q_{\alpha \beta} = \sum_i e_{i \alpha} e_{i \beta} (f_i - f_i^{\mathrm{eq}})$:
\begin{equation}
	\tau_t = \frac{1}{2} \left( \sqrt{\tau_0^2 + 4 c_s^{-4} C_S^2 (Q_{\alpha \beta} Q_{\alpha \beta})^{1/2}} - \tau_0 \right)
	\label{eq:taut}
\end{equation}
where $C_S$ is the Smagorinsky constant, which for LB models can be taken to be $0.1$.
This effectively abandons the single relaxation time approximation by making it a spatially and
temporally varying quantity dependent on the local gradients of the fluid velocity \cite{yu_dns_2005}.

\subsection{Multi-fluid models}

The multi-fluid models discussed here are all diffuse interface models, without any explicit
interface tracking -- the dynamics of the interface emerges naturally from the interactions
between the fluid components. All models in this class use an additional lattice (whose mass
fractions we will call $g_i$) to represent the second fluid species and nonlocal interactions
between the lattices in the collision process.

\subsubsection{Shan-Chen}

In the Shan-Chen model, both lattices are used to represent fluid components, with
the 0-th moment of $f_i$ and $g_i$ representing the density of the first (A) and
second fluid components (B), respectively.  The equilibrium function and relaxation
schemes remain unchanged for both lattices, but an additional coupling term in the form of a body force:
\begin{equation}
	\vec{F}_A(\vec{x}) = G \psi_A(\vec{x}) \sum_{i} w_i \psi_B(\vec{x} + \vec{e}_i) \vec{e}_i
	\label{eq:sc_force}
\end{equation}
is introduced into \eqref{eq:lbgk} (see Section~\ref{sec:body_forces} for a short overview of ways of adding
body forces to an LB simulation), where $\psi_B(\vec{x})$ is a pseudopotential function,
dependent on the density $\rho_B$ at node $\vec{x}$ and $G$ is a coupling constant. A similar
term $\vec{F}_B$ of the form \eqref{eq:sc_force} with $\psi_A$ replaced by $\psi_B$ and vice versa
 is added to the collision operator for the second component.

A commonly used pseudopotential function is
$\psi(\rho) = 1 - e^{-\rho}$. The velocity of the fluid becomes a weighted average of the
first moments of the distribution functions:
$\vec{u} = \frac{\rho_A \vec{u}_A / \tau_A + \rho_B \vec{u}_B / \tau_B}{\rho_A / \tau_A + \rho_B / \tau_B}$,
where $\vec{u}_A$, $\vec{u}_B$ and $\rho_A$, $\rho_B$ are velocities and densities
computed respectively from $f_i$ and $g_i$ using \eqref{eq:lb_macro},
and $\tau_A$, $\tau_B$ are relaxation times for the two fluid components.

\subsubsection{The free energy model}

The free energy model is based on a Landau free energy functional
for a binary fluid \cite{PhysRevE.78.056709}. In the LB realization,
the 0-th moment of $f_i$ represents the density $\rho$ of both components,
while the 0-th moment of $g_i$ is a smoothly varying order parameter $\phi$, with $-1$ indicating
a pure first component and $1$ indicating a pure second component:
\begin{equation}
	\rho = \sum_i f_i \qquad \phi = \sum_i g_i \qquad \rho \vec{u} = \sum_i \vec{e}_i f_i.
	\label{eq:free_energy_macro}
\end{equation}
The macroscopic dynamics of the fluid is described by the system of equations
\begin{align}
\begin{split}
	\partial_t \rho + \partial_{\alpha} (\rho u_{\alpha}) &= 0 \\
	\partial_t (\rho u_{\alpha}) + \partial_{\alpha}(\rho u_{\alpha} u_{\beta}) &= -\partial_{\alpha} P_{\alpha \beta} + \partial_{\alpha} \{ \nu \rho (\partial_{\beta} u_{\alpha} + \partial_{\alpha} u_{\beta})\} \\
	\partial_t \phi + \partial_{\alpha} (\phi u_{\alpha}) &= M \nabla^2 \mu
	\label{eq:free_energy_dynamics}
\end{split}
\end{align}
where $M$ is a mobility parameter, $\mu$ is the chemical potential and
the pressure tensor $P_{\alpha \beta}$ is defined as:
\begin{equation}
	P_{\alpha \beta} = \left(p_0 - \kappa \phi \nabla^2 \phi - \frac{\kappa}{2} |\nabla \phi |^2 \right) \delta_{\alpha \beta} + \kappa \partial_{\alpha} \phi \partial_{\beta} \phi,
\end{equation}
and $p_0 = \rho/3 + a\left(\phi^2 / 2 + 3 \phi^4/4 \right)$ is the bulk pressure.

In the corresponding LB scheme, distributions on both lattices are relaxed and streamed using the standard LBGK
scheme \eqref{eq:lbgk} using the relaxation times $\tau_\rho = \tau_B + \frac{\phi + 1}{2} (\tau_A - \tau_B)$ and $\tau_\phi$, respectively.
The following equilibrium functions \cite{PhysRevE.78.056709} are used to recover
\eqref{eq:free_energy_dynamics} in the macroscopic limit:
\begin{align}
	f_i^{\mathrm{eq}} &= w_i \left(p_0 - \kappa \phi \nabla^2 \phi + e_{i \alpha} u_\alpha \rho +
			\frac{3}{2} \left[ e_{i \alpha} e_{i \beta} - \frac{\delta_{\alpha \beta}}{3} \right] \rho u_{\alpha} u_{\beta}
			\right) + \kappa w_i^{\alpha \beta} \partial_{\alpha} \phi \partial_{\beta} \phi \\
	g_i^{\mathrm{eq}} &= w_i \left( \Gamma \mu + e_{i \alpha} u_{\alpha} \phi +
			\frac{3}{2} \left[ e_{i \alpha} e_{i \beta} - \frac{\delta_{\alpha \beta}}{3} \right] \phi u_{\alpha} u_{\beta} \right)
	\label{eq:free_energy_eq}
\end{align}
where $\Gamma$ is a tunable parameter related to mobility via $M = \Gamma (\tau_\phi - 1/2)$,
and the chemical potential $\mu = a( -\phi + \phi^3) -\kappa \nabla^2 \phi$. $\kappa$ and $a$
are constant parameters related to the surface tension $\gamma = \sqrt{8\kappa a / 9}$ and interface
width $\xi = 2 \sqrt{2\kappa/a}$. The values of the weights $w_i$ and $w_i^{\alpha \beta}$
can be found in the paper by Kusumaatmaja and Yeomans~\cite{kusumaatmaja_contact_2008}.

In order to minimize spurious currents at the interface between the two fluids, we use optimized
gradient and Laplacian stencils \cite{PhysRevE.77.046702}.

\subsection{Boundary conditions}

Sailfish implements various boundary conditions, which taken together make it possible
to model a wide range of physical situations:
\begin{itemize}
	\item periodic boundary conditions (used to simulate a system spatially infinite along a given axis),
	\item no-slip (solid walls): half-way bounce-back, full-way bounce-back (both with configurable wetting when used with the free energy model),
       Tamm-Moth-Smith~\cite{chikatamarla_entropic_2013},
	\item density/pressure: Guo's method, Zou-He's method, equilibrium distribution, regularized~\cite{PhysRevE.77.056703},
	\item velocity: Zou-He's method, equilibrium distribution, regularized~\cite{PhysRevE.77.056703},
	\item outflow: Grad's approximation~\cite{grad_outflow}, Yu's method~\cite{yu2005improved}, Neumann's, nearest-neighbor copy, ,,do nothing''~\cite{junk2008}.
\end{itemize}

\begin{figure}
	\centering
	\includegraphics[width=\textwidth]{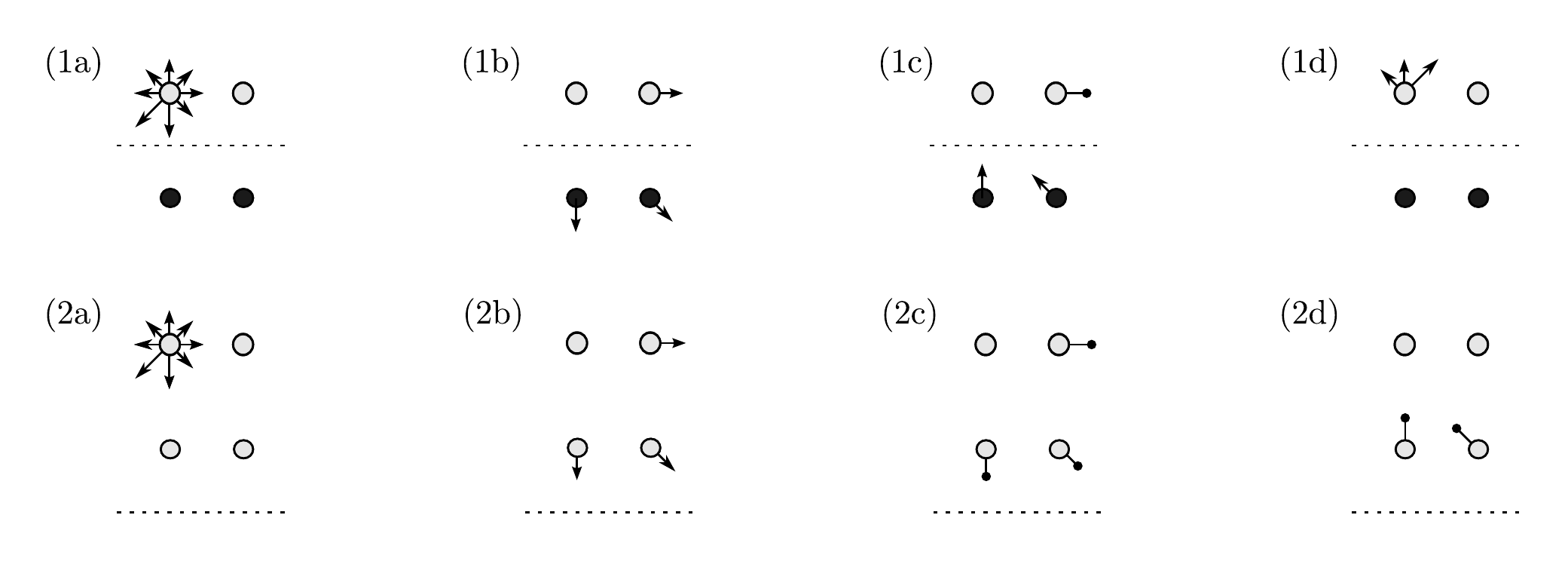}
	\label{fig:bb}
	\caption{The bounce-back scheme illustrated for the D2Q9 lattice. (1) full-way bounce-back, (2) half-way bounce-back. The bottom row of
nodes within each group represents the bounce-back nodes, while the top row represents normal fluid. Black nodes are ,,dry'' (do not represent fluid, do not undergo relaxation),
while grey nodes are ,,wet''. The dashed line represents the effective location of the no-slip boundary condition. For simplicity, only distributions originating from the top left
node are shown in subsequent time steps. The steps are: (a) initial state, time $t$ (b) streaming, time $t+1$ (c) relaxation, time $t+1$ (arrows ending with disks represent the
post-relaxation state), (d) streaming, time $t+2$.}
\end{figure}

The bounce-back method is a simple idea originating from lattice gas automatons -- a particle distribution
is streamed to the wall node, and scattered back to the node it came from (see \figref{bb}). In the full-way bounce-back scheme
the wall nodes do not represent any fluid (,,dry'' nodes) and are only used to temporarily store distributions
before scattering. This approach is slightly easier to implement, but has the downside of the distributions needing
two time steps to reach back to the originating fluid node (one step to reach the wall node, then another to
be reflected back and streamed to the fluid node). The physical wall is effectively located between the last fluid
node and the wall node. In contrast, in the half-way bounce-back the wall nodes do represent fluid, the wall is located
beyond the wall node (away from the fluid domain), and the distributions are reflected in the same time step in which
they reach the wall node~\cite{Ladd94A}.

For other types of boundary conditions, the problem is more complex, as the mass fractions pointing into the
fluid domain are undefined (there are no nodes streaming to them in these directions). Various schemes have
been proposed to work around this problem. In the equilibrium approach, the missing incoming distributions are
assumed to have the same values as those opposite to them (bounce-back), and the post-collision distributions
are defined to be $f_i^{\mathrm{eq}}(\rho, \vec{u})$ with one of $\{\rho, \vec{u}\}$ being specified
as part of the boundary condition, and the other calculated from the local distribution function.
Zou's and He's method \cite{zouhe} uses the idea of the bounce-back
of the non-equilibrium parts of the distributions to calculate the missing mass fractions.
The regularized boundary conditions also use the non-equilibrium bounce-back idea, but only
for an intermediate step to calculate the 2-nd moment of the non-equilibrium distribution,
which is then used to reset all mass fractions at the boundary node via a regularization
procedure. We refer the reader to the excellent paper by Latt et al.~\cite{PhysRevE.77.056703} for a
detailed overview of these and other boundary conditions.

The Grad's approximation method uses a formula dependent on
the density, velocity and the pressure tensor $P_{\alpha \beta} = \sum_i e_{i\alpha} e_{i \beta} f_i$
to replace all distributions on the boundary node. The three macroscopic quantities can be taken
to be those from the boundary node at the previous time step (less precise, fully local)
or extrapolated from neighboring fluid nodes (more precise, nonlocal). Other outflow
methods, such as the Neumann's boundary implementation or Yu's method, are also nonlocal
and use a simple extrapolation scheme. The simplest methods (nearest-neighbor copy, do nothing)
do not require extrapolation, but also give less control over the resulting macroscopic fields.

\subsection{Body forces}
\label{sec:body_forces}

Sailfish implements two popular ways of adding a body force $\vec{F}$ to the simulation: Guo's method \cite{GuoForce},
and the Exact Difference Method (EDM) \cite{Kupershtokh2009965}.
In both schemes, the actual fluid velocity $\vec{v}$ can be computed as
\begin{equation}
	\rho \vec{v} = \sum_i \vec{e}_i f_i + \vec{F} / 2 = \rho \vec{u} + \vec{F} / 2,
	\label{eq:edm_momentum}
\end{equation}
and a force term $F_i$ is added to the right hand side of \eqref{eq:lbgk}. Guo et al. analyzed
a number of popular schemes and concluded that the optimal choice for $F_i$ in an LBGK simulation is
\begin{equation}
	F_i = \left(1 - \frac{1}{2\tau} \right) w_i
		\left[ \frac{\vec{e}_i - \vec{v}}{c_s^2} + \frac{\vec{e}_i \cdot \vec{v}}{c_s^4} \vec{e}_i  \right] \cdot \vec{F}.
	\label{eq:guo_force}
\end{equation}

In the EDM, the body force term added to the collision operator is equal to the difference of equilibrium
distribution functions computed using momentum after and before the action of the force $\vec{F}$:
\begin{equation}
	F_i = f_i^{\mathrm{eq}}(\rho, \vec{u} + \Delta\vec{u}) - f_i^{\mathrm{eq}}(\rho, \vec{u})
	\label{eq:edm_force}
\end{equation}
with $\Delta\vec{u} = \vec{F}/\rho$. The advantages of the EDM are its lack of any spurious terms
in the macroscopic Navier-Stokes equations and its
applicability to any collision operator (not only LBGK).

\section{Software design and algorithms}

Computational software is traditionally developed in one of two different paradigms.
\textit{Prototype} code is written with relatively little effort
in a high level environment such as Matlab or Python for exploratory or teaching purposes.
Due to the relatively low performance of these environments, this type of code is unsuitable for
large scale problems.  In contrast, \textit{production} code is written in lower level
languages such as C++ or Fortran in a higher effort process, which can sometimes span
many years and involve teams of developers.  The resulting programs are more efficient, but also longer, less
readable, and potentially difficult to maintain.

In our implementation of the lattice Boltzmann method for GPUs released under the name
\textit{project Sailfish}, we took a hybrid approach. We use the Python programming language
with a template-based code generation module (using the Mako language) and a computer algebra system (Sympy) to
generate code in CUDA C or OpenCL. We chose Python because it is a very expressive
language, with bindings to many system libraries and great support for GPU programming
via the PyCUDA and PyOpenCL packages~\cite{klockner2012pycuda}. In Sailfish,
Python is used for setting up the simulation (initial
conditions, boundary conditions, selecting an LB model), for simulation control, communication
(e.g. between compute nodes in a cluster) and input/output (loading geometry, saving simulation results).
We also employ the NumPy package to perform matrix operations efficiently.

\begin{lstlisting}[label=lst:equilibrium,language=Python,caption=BGK equilibrium function defined using Sympy expressions. \texttt{S} is a class with predefined symbols. \texttt{grid} is a class representing the lattice used for the simulation. Note that this code works for all lattices and dimensions. The discrete velocity vectors are stored in symbolic form in the \texttt{grid.basis} list. This allows for e.g. simple computation of dot products. Numerical coefficients in the formulas are also stored in symbolic form as rational expressions (e.g. \texttt{Rational(3, 2)}) instead of floating point values. This makes them exact and allows further symbolic simplification.]
from sympy import Rational
def bgk_equilibrium(grid):
    out = []
    for ei, weight in zip(grid.basis, grid.weights):
        out.append(weight * (S.rho + S.rho * (3 * ei.dot(grid.v) + Rational(9, 2) * (ei.dot(grid.v))**2 - Rational(3, 2) * grid.v.dot(grid.v))))

    return out
\end{lstlisting}

Mathematical
formulas, such as \eqref{eq:lb_macro}, \eqref{eq:equilibrium}, \eqref{eq:free_energy_eq}
are stored in the form of Sympy expressions. We decided to do this after noticing that
computer code for numerical calculations is often very repetitive, e.g. \eqref{eq:equilibrium},
when evaluated on a D3Q19 lattice, would expand to 19 long lines of code, even though
the initial formula easily fits in one line (see \lstref{equilibrium}).
This approach makes the code easier to read (the formulas can
also be automatically converted to LaTeX expressions), allows for automated consistency
checks (e.g. one can easily, in a single line of code, verify that the 0th
moment of $\feq$ is $\rho$), enables easy experimentation with code optimization
techniques (formulas can be reorganized e.g. for speed or precision of floating-point operations),
and enables code reuse (e.g. the formula for equilibrium or the bounce-back rule is written
down once, but can generate code for various types of lattices, both in 2D and 3D, see \lstref{bb} for an example).

Finally, the Mako template language is used to render the formulas into low-level CUDA/OpenCL code.
The generated code is fairly small, contains comments, and is automatically formatted,
making it suitable for instructional purposes as well as for one-off, ad-hoc modifications.
This is in stark contrast to large, multimodular codes where the architecture of the
system can be difficult to understand without extensive documentation.

Sailfish was designed from scratch and optimized for modern massively parallel architectures,
such as GPUs. We wanted to achieve high flexibility of the code, ease of use when running or defining
new simulations, and not take any performance hits compared to code written directly
in CUDA C or OpenCL. In order to achieve that, we decided to use run-time code generation techniques.
This provides some isolation from low-level hardware details, important in the case of GPUs
which are still evolving rapidly and changing with every new GPU architecture. It also makes
it possible to generate optimized code on a case-by-case basis and to automatically
explore parameter spaces to find optimal solutions, thus saving programmer time and
increasing their productivity. With many details, such as the number of compute units,
size of on-chip memory, speed of access patterns to on-chip and off-chip memory,
 host-device latency and bandwidth, memory bandwidth to computational performance ratio directly
impacting the performance of the code, experimentation and microbenchmarking are necessary
to find combinations of parameters that work well.

We also considered other metaprogramming approaches to code generation, such as domain-specific languages,
and templates in C++.
We deemed the first solution to have too
much overhead, and decided against the latter one since the expanded code cannot be saved
for inspection and modification, and there were no open source computer algebra libraries providing
a level of flexibility and sophistication comparable to Sympy.

\begin{lstlisting}[multicols=2,label=lst:bb,escapechar=\!,language=C,morekeywords={endfor,in,\$},caption=Full-way bounce-back rule. Left column: source code in Mako. Right column: generated CUDA C code for the D2Q9 lattice.]

${device_func} inline void bounce_back(Dist *fi)

	float t;

	%for i in sym.bb_swap_pairs(grid):
		<% a = grid.idx_name[i]
		   opp_i = grid.idx_opposite[i]
		   b = grid.idx_name[opp_i] %>
		t = fi->${a};
		fi->${a} = fi->${b};
		fi->${b} = t;
	%endfor
}
!\vfill\columnbreak!
__device__ inline void bounce_back(Dist * fi)
{
	float t;
	t = fi->fE;
	fi->fE = fi->fW;
	fi->fW = t;
	t = fi->fN;
	fi->fN = fi->fS;
	fi->fS = t;
	t = fi->fNE;
	fi->fNE = fi->fSW;
	fi->fSW = t;
	t = fi->fNW;
	fi->fNW = fi->fSE;
	fi->fSE = t;
}


\end{lstlisting}

\subsection{High-level simulation architecture}

Sailfish takes extensive advantage of objected-oriented programming techniques for modularization
and code reuse purposes. Each simulation is defined using two classes -- a simulation class,
and a geometry class (see \lstref{ldc_example}). The simulation domain can be divided into cuboid subdomains, which do
not need to fill the parts of the domain that do not contain any fluid (this makes it
possible to handle complex geometries). The geometry class defines initial and boundary conditions
for a single subdomain. The simulation class derives from a base class specific to the LB
model used such as \texttt{LBFluidSim} (single component simulations) or
\texttt{LBBinaryFluidFreeEnergy} (binary fluids using the free energy model). The simulation
class can optionally also add body forces or define custom code to be run after selected
steps of the simulation (e.g. to display a status update, check whether the steady state
has been reached, etc).

The base simulation class specifies the details of the used LB model, such as the
form of the equilibrium distribution function (in symbolic form, stored as a Sympy expression),
number and names of macroscopic fields (density, velocity, order parameter, etc), and names
of GPU code functions that need to be called at every simulation step.

\begin{figure}
	\centering
	\includegraphics[width=0.9\textwidth]{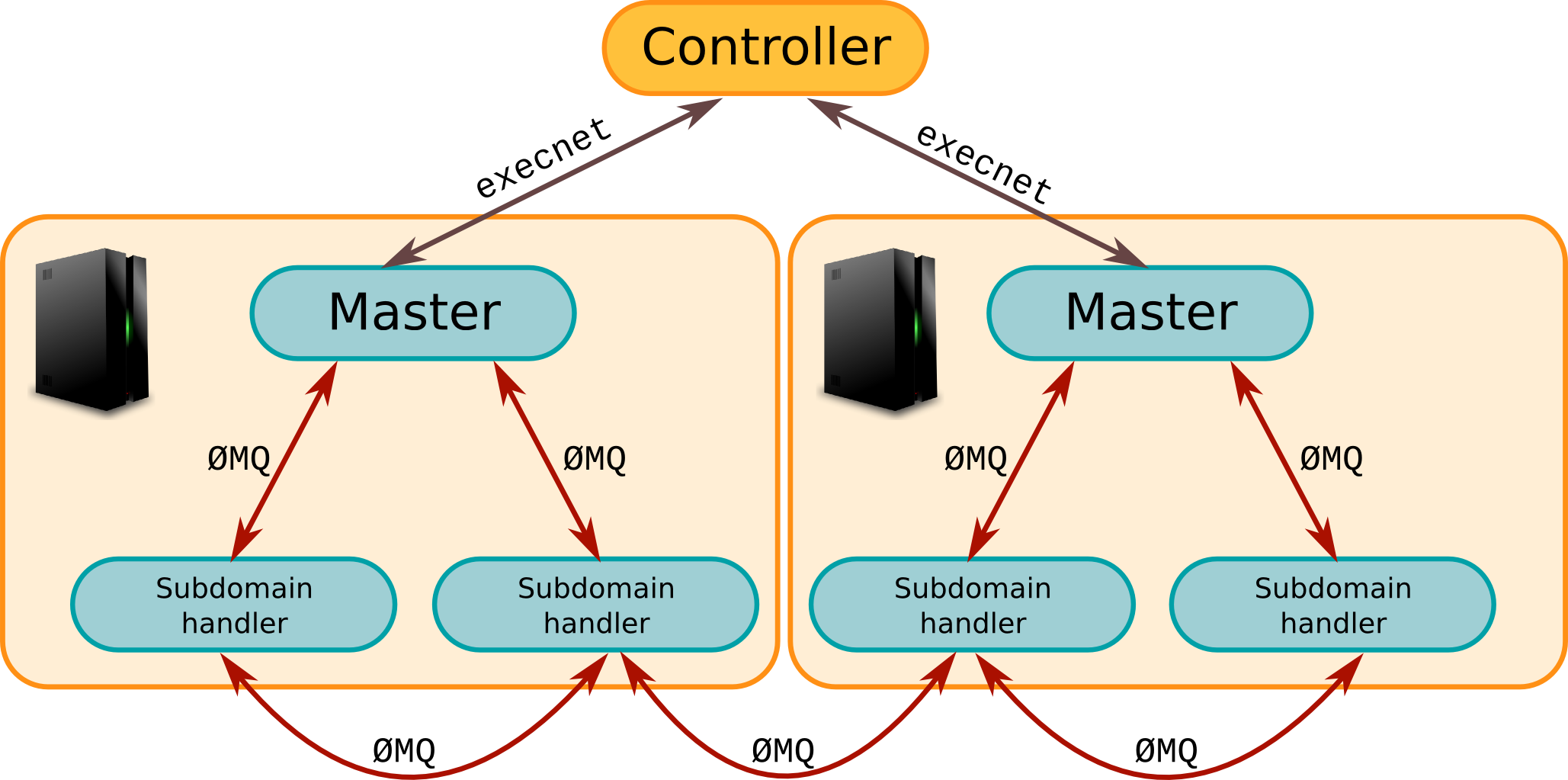}
	\caption{High-level architecture of a distributed Sailfish simulation. The simulation
is divided into 4 subdomains. The controller process uses the execnet Python library to start
machine master processes on two computational nodes, which then spawn children processes
to handle individual subdomains. The master and subdomain handlers use the ZeroMQ library
to communicate.}
	\label{fig:system_arch}
\end{figure}

When executed, every Sailfish simulation starts a controller process which parses any
command line parameters and reads configuration files, decides how many computational nodes
and GPUs are to be used (in case of distributed simulations), and how the subdomains are going
to be assigned to them. It then starts a master process on every computational node
(see \figref{system_arch}), which in turn starts a subdomain handler process for every
subdomain assigned to the computational node. We use subprocesses instead of threads in
order to avoid limitations of the Python runtime in this area (the global interpreter
lock preventing true multithreading), as well as to simplify GPU programming. The subdomain
handlers instantiate the simulation and geometry classes for their respective subdomains.
The system state is initialized in the form of Numpy arrays describing the initial
macroscopic fields, and these are then used to initialize the distribution functions
on the GPU (using the equilibrium distribution function). Optionally, a self-consistent
initialization procedure \cite{luo-initial} can be performed to compute the density
field from a known velocity field. Once the system is fully initialized, the
simulation starts running, with the subdomain handlers exchanging information in a peer-to-peer
fashion as determined by the connectivity of the global geometry.

\begin{lstlisting}[label=lst:ldc_example,language=Python,caption=2D Lid-driven cavity example in Sailfish.]
from sailfish.subdomain import Subdomain2D
from sailfish.node_type import NTFullBBWall, NTZouHeVelocity
from sailfish.controller import LBSimulationController
from sailfish.lb_single import LBFluidSim

class LDCSubdomain(Subdomain2D):
    max_v = 0.1

    def boundary_conditions(self, hx, hy):
        wall_map = (hx == self.gx-1) | (hx == 0) | (hy == 0)
        self.set_node((hy == self.gy-1) & (hx > 0) & (hx < self.gx-1), NTZouHeVelocity((self.max_v, 0.0)))
        self.set_node(wall_map, NTFullBBWall)

    def initial_conditions(self, sim, hx, hy):
        sim.rho[:] = 1.0
        sim.vx[hy == self.gy-1] = self.max_v

class LDCSim(LBFluidSim):
    subdomain = LDCSubdomain

if __name__ == '__main__':
    LBSimulationController(LDCSim).run()
\end{lstlisting}

\subsection{GPU architecture overview}

Modern GPUs are massively parallel computational devices, capable of performing trillions of floating-point
operations per second. We will now briefly present the architecture of CUDA-compatible devices
as a representative example of the hardware architecture targeted by Sailfish.
Other devices not supporting CUDA but supporting OpenCL are based
on the same core concepts. CUDA devices can be grouped into three generations, called
Tesla, Fermi, and Kepler -- each one offering progressively more advanced features and better performance.

The CUDA GPU is organized around the concept of a streaming multiprocessor (MP).
Such a multiprocessor consists of several scalar processors (SPs), each of which is
capable of executing a thread in a SIMT (Single Instruction, Multiple Threads) manner.
Each MP also has a limited amount of specialized on-chip memory: a set of 32-bit registers,
a shared memory block and L1 cache, a constant cache, and a texture cache. The registers are logically local
to the scalar processor, but the other types of memory are shared between all SPs
in a MP, which allows data sharing between threads.

Perhaps the most salient feature of the CUDA architecture is the memory hierarchy
with 1-2 orders of magnitude differences between access times at each successive level.
The slowest kind of memory is the host memory (RAM). While the
RAM can nowadays be quite large, it is separated from the GPU by the PCIe bus, with a
maximum theoretical throughput in one direction of 16 GB/s (PCI Express 3.0, x16 link).

Next in line is the global device memory of the GPU, which is currently limited to several
gigabytes and which has a bandwidth of about 100-200~GB/s. Global memory accesses are however
high-latency operations, taking several hundred clock cycles of the GPU to complete.

The fastest kind of memory currently available on GPUs is the shared memory block residing
on MPs. It is currently limited in size to just 48~kB (16~kB on Tesla devices), but has
a bandwidth of ca~1.3~TB/s and a latency usually no higher than that of a~SP register access.

The above description readily suggests an optimization strategy which we will generally
follow in the next section and which can be summarized as: move as much data as possible
to the fastest kind of memory available and keep it there as long as possible, while minimizing
accesses to slower kinds of memory. When memory accesses are necessary, it also makes
sense to try to overlap them with independent computation, which can then be executed
in parallel effectively hiding the memory latency.

From the programmer's point of view, CUDA programs are organized into kernels. A kernel
is a function that is executed multiple times simultaneously on different MPs. Each
instance of this function is called a thread, and is assigned to a single scalar processor.
Threads are then grouped in one-, two- or three-dimensional blocks assigned to multiprocessors
in an 1-1 manner (1 block –- 1 MP). The blocks are organized into a one- or
two-dimensional grid. The size and dimensionality of the grid and blocks is determined
by the programmer at the time of kernel invocation. Knowledge of the grid position, and
the in-block position makes it possible to calculate a thread ID that is unique during
the kernel execution. Within a single block threads can synchronize their execution and
share information through the on-chip shared memory. Synchronization between blocks
is not supported in any way other than serializing the execution of kernels, and through
atomic operations on global memory.

\subsection{Low-level LB algorithms and data structures on GPUs}

The great potential of modern GPUs as a hardware platform for simulating
both two-dimensional \cite{tolke-cuda-twod} and three-dimensional flows \cite{tolke-GPU}
with the lattice Boltzmann method was quickly realized after the initial release of the CUDA programming
environment in 2007. Recently, GPUs have also been used to investigate more complex,
two-dimensional multicomponent flows \cite{PhysRevE.80.066707}.

There are three main factors that need to be carefully optimized in order to fully utilize
the computational power of modern GPUs: memory access patterns, register utilization, and overlap
of memory transfers and arithmetic operations.  The first factor has received the most
attention in the literature, but in this work we show that the remaining two are also important.

The most important data structure used in an LB simulation are the particle distributions $f_i$.
They are stored in the global GPU memory in a Structure of Arrays (SoA) fashion, effectively
forming a 4D or 3D array (for 3D and 2D simulations, respectively), with the following index ordering:
$(q, z, y, x)$, where $q$ is the number of discrete velocities in the lattice. An
Array of Structures (AoS) approach, while elegant from the point of view of object-oriented programming,
is completely unsuitable for GPU architectures due to how global reads and writes are performed by the
hardware.

Global memory accesses are performed by thread warps (32 threads)
in transactions of 32, 64, or 128 bytes. On Fermi devices, accesses cached in the L1 cache
are serviced with 128-byte transactions (the size of a full L1 cache line), while those
cached in the L2 cache are serviced with 32-byte transactions.
In order to attain good bandwidth utilization, the memory has to be accessed in contiguous
blocks so that all bytes in a transaction are used for meaningful data. The memory location
also has to be naturally aligned, i.e. the first address in the transferred segment
must by a multiple of the segment's size.


In Sailfish, we run all kernels in 1-dimensional thread blocks spanning the X axis of the subdomain,
with a typical block size being 64, 128 or 192 nodes. Each thread handles a single node of the
subdomain. Due to the layout of the distributions
in global memory, we issue $q$ read requests to load a full set of mass fractions in a thread
block. Each request results in a fully utilized memory transaction. To ensure natural alignment
of transactions, the X dimension of the distributions array in global memory is padded with
unused nodes so that $x$ is a multiple of 32 or 16, for single and double precision floating
point numbers, respectively.

In addition to the distributions array, we also store a node type map (see Section~\ref{sec:boundary}),
and macroscopic field arrays in global memory, all following the same memory layout and padding
as described above.

A simple LB simulation in Sailfish repeatedly calls a single CUDA kernel called \texttt{CollideAndPropagate},
which implements both the collision and propagation step (see e.g. \eqref{eq:lbgk}), as well as any
boundary conditions. A high-level summary of this kernel is presented in \algref{cnp}.

\begin{algorithm*}
	\caption{A GPU kernel to perform one step of a lattice Boltzmann simulation, using shared memory
		for propagation in the $x$ direction. $\mathrm{offset}(x, y, z)$ is a
		function computing a linear index in a 1D array corresponding to a shift in the 3D subdomain space.
		$i$ is a thread index within the local block of threads (\texttt{threadIdx.x} in CUDA C). $e_{km}$
		is the $m$-th component of the $k$-th discrete velocity vector.}
	\label{alg:cnp}
	\singlespacing
	\begin{algorithmic}[1]
		\State allocate shared memory array $\mathit{buf}_k[\mathit{blockSize}]$
		\State compute global array index $i_G$ for the current node $i$
		\State load and decode node type from global memory to $nt$
		\State load distributions from global memory to $f$
		\State compute macroscopic variables $\rho$ and $\vec{v}$
		\If{$nt$ is a boundary node}
			\State apply boundary conditions
		\EndIf
		\If{output requested}
			\State save $\rho$ and $\vec{v}$ in global memory arrays
		\EndIf
		\State compute the equilibrium distribution $\feq$ using $\rho$ and $\vec{v}$
		\State relaxation: $f \gets f + \frac{1}{\tau} (\feq - f)$
		\For{$k$ such that $e_{k x} = 0$}
			\Comment{Propagation in directions orthogonal to X.}
			\State write $f_k$ to global memory at $i_G + \mathrm{offset}(0, e_{k y}, e_{k z})$
		\EndFor
		\If{$i \ge 1$}
			\Comment{Propagation backward in the X direction.}
			\For{$k$ such that $e_{k x} = -1$}
				\State $\mathit{buf}_k[i-1] \gets f_{k}$
			\EndFor
		\Else
			\For{$k$ such that $e_{k x} = -1$}
				\State write $f_k$ to global memory at $i_G + \mathrm{offset}(e_{k x}, e_{k y}, e_{k z})$
			\EndFor
		\EndIf
		\State synchronize threads within the block
		\If{$i < \mathit{blockSize} - 1$}
			\For{$k$ such that $e_{k x} = -1$}
				\State write $\mathit{buf}_k[i]$ to global memory at $i_G + \mathrm{offset}(0, e_{k y}, e_{k z})$
			\EndFor
			\For{$k$ such that $e_{k x} = 1$}
				\Comment{Propagation forward in the X direction.}
				\State $\mathit{buf}_k[i+1] \gets f_{k}$
			\EndFor
		\Else
			\For{$k$ such that $e_{k x} = 1$}
				\State write $f_k$ to global memory at $i_G + \mathrm{offset}(e_{k x}, e_{k y}, e_{k z})$
			\EndFor
		\EndIf
		\State synchronize threads within the block
		\If{$i > 0$}
			\For{$k$ such that $e_{k x} = 1$}
				\State write $\mathit{buf}_k[i]$ to global memory at $i_G + \mathrm{offset}(0, e_{k y}, e_{k z})$
			\EndFor
		\EndIf
	\end{algorithmic}
\end{algorithm*}

\begin{figure}
	\centering
	\includegraphics{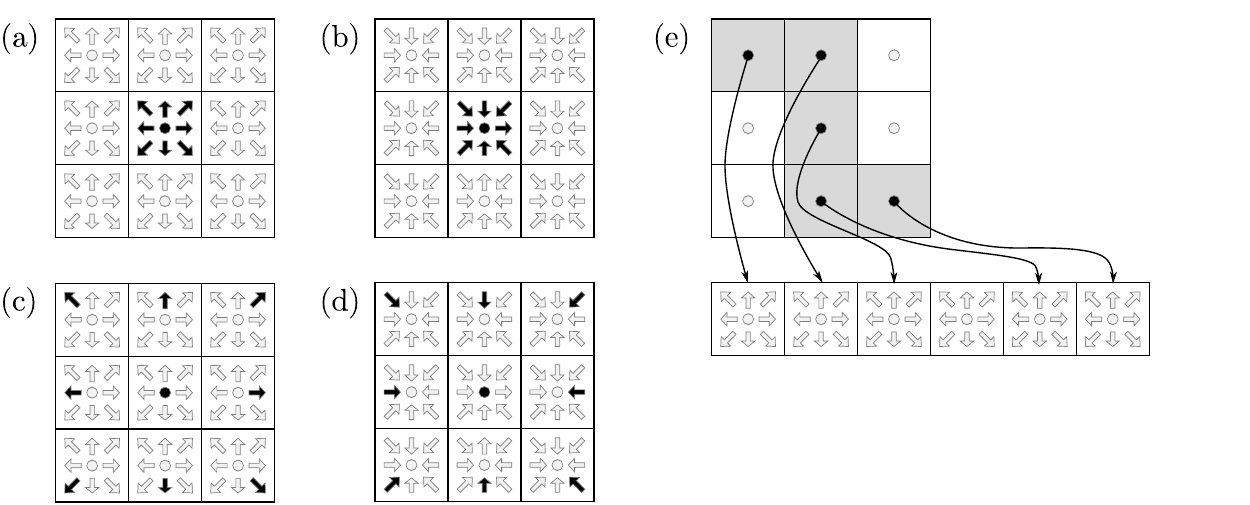}
	\label{fig:memory}
	\caption{Memory layouts supported by Sailfish. In (a)-(d), dark arrows represent distributions used for calculations on the central node (middle square in every panel). In the AB layout, there are two copies of the lattice stored in memory. Data is read from the first lattice (a), and propagated into the second lattice (c).
In the AA layout, there is only one lattice copy. Data is read from (a), stored as (b), and in the following step, read from (d) and stored as (c). In indirect addressing, there is a dense array of
pointers to the actual distributions, which are stored in a continuous array (e). Dark nodes with black circles represent active (fluid) nodes. White nodes represent areas outside of the simulation
domain which do not have corresponding distributions.}
\end{figure}

Two basic patterns of accessing the distributions have been described in the literature, commonly
called AB and AA~\cite{bailey_2009}. In the AB
access pattern, there are two copies of the distributions in global memory (A and B). The simulation
step alternates between reading from A and writing to B, and vice versa.
In the AA access pattern, only a single copy of the distributions array is stored in global memory.
Since there are no guarantees about the order of execution of individual threads,
care has to be taken that a thread reads from and writes to exactly the same locations in memory.
This is illustrated in~\figref{memory}. A third access pattern that is used in practice is
the so-called \emph{indirect addressing}. In this mode, in order to access data for a node, its
address has to be read first. This causes overhead both in storage (need to store the addresses)
and in memory accesses, but can be very useful for complex geometries where the active nodes
are only a tiny fraction of the volume of the bounding box. Indirect addressing can be combined
with both AA and AB access patterns.  For a dense subdomain using indirect
addressing, the performance can be 5-25\% lower than when using direct addressing, with C2050
showing better results with the AA access pattern than in AB, and the GTX 680 exhibiting
the opposite tendency. The exact performance is however necessarily geometry-dependent, and as such
it is not discussed further here.

The propagation step of the LB algorithm shifts the distributions by $\pm 1$ node in all directions
-- when this happens for the X axis, it results in misaligned reads/writes.
To reduce the impact of misaligned writes, Sailfish utilizes shared memory to shift data in the X
direction within the thread block (see~\algref{cnp}).

We performed numerical tests
to evaluate the impact of various data access patterns on the overall
performance of the simulation.  Earlier works \cite{Obrecht2010,ObrechtMemory} indicate that the cost of unaligned
writes is significantly higher than the cost of unaligned reads, and that a propagate-on-read
strategy results in up to 15\% performance gain.  Our experiments confirmed this on older GT200
hardware (GTX 285), however we failed to replicate this effect on Fermi and Kepler devices (Tesla C2050, K10, K20),
where the performance of both implementations was nearly identical (typically, with a few \% loss for propagate-on-read).
This is most likely caused by hardware improvements in the Fermi and later architectures.

We also compared the performance for the AB and AA access patterns. On Tesla-generation devices (GTX 285, Tesla C1060),
the AA memory layout results in a slightly higher performance, and is therefore clearly preferred. On newer devices the 
AB scheme is typically a little faster, but the performance gains over the AA scheme are minor ($<10\%$), and as such the AB
scheme is only preferable when ample GPU memory is available.

\subsection{Boundary condition handling}
\label{sec:boundary}

\begin{figure}
	\centering
	\includegraphics[width=\textwidth]{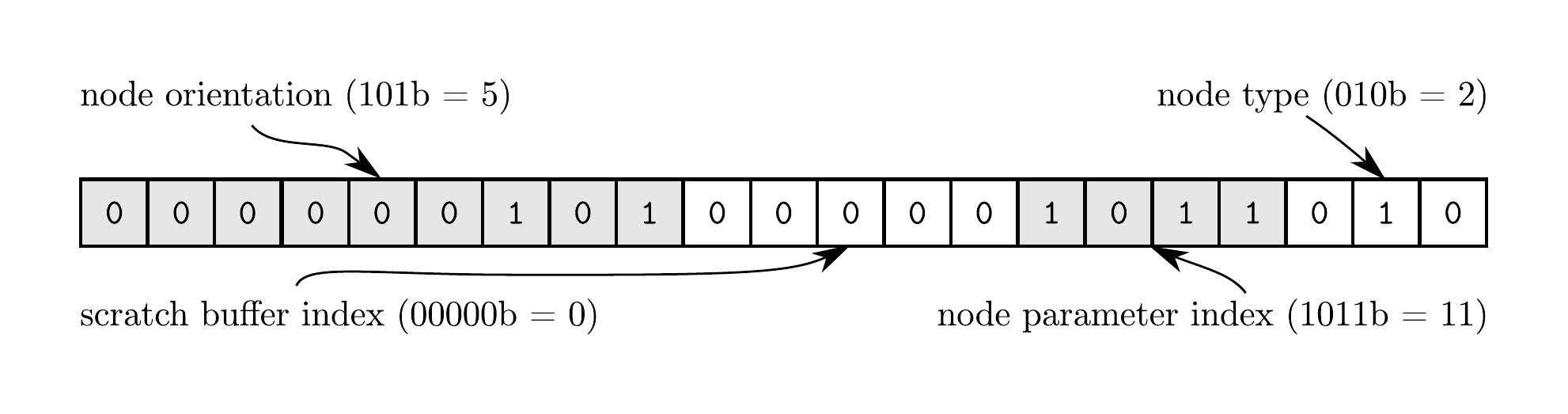}
	\caption{Node parameter and type encoding into a single 32-bit unsigned integer to be used in the node type map.
Bitfields (node type, node parameter index, ...) have adjustable size which can vary between simulations.}
	\label{fig:encoding}
\end{figure}
Boundary conditions are handled in Sailfish with the help of a \emph{node type map} -- an unsigned
32-bit integer array stored in the global GPU memory. Each entry in this array contains encoded
information about the node type (fluid, unused, ghost, type of boundary condition to use),
orientation (vector normal to the boundary, pointing into the fluid) and a parameter ID.
The parameter ID is an index to a global array of values used by boundary conditions (e.g.
densities, velocities). The encoding scheme uses variable-size bitfields, which are dynamically
chosen for every simulation depending on the usage of different boundary conditions in a subdomain
(see \figref{encoding}).

Time-dependence is supported for all types of boundary conditions. When a value changing in time
is required, it is typically specified in the form of a Sympy expression. This expression
is then transformed into a GPU function and assigned an ID, analogous to the parameter ID
for static boundary conditions.


\subsection{Multicomponent models}

Models with more than one distribution function, such as the Shan-Chen model or the free energy
model introduce a nonlocal coupling via one or more macroscopic fields. To minimize the amount
of data read from global memory, we split the simulation step into two logical parts, implemented
as two GPU kernels. In the first kernel, \texttt{ComputeMacroFields}, we load the distributions
and compute the macroscopic field value for every field which needs to be accessed in a nonlocal
manner in the collision kernel. The collision step is implemented similarly to single fluid
models, and the nonlocal quantities (Shan-Chen force, gradient and Laplacian of the order parameter
in the free energy model) are computed by directly accessing the macroscopic field values in global memory.

We considered three approaches to realizing the collision and propagation part of the algorithm
for multifluid models. In the first variant, we used a single collision kernel, which loaded
both distributions into registers and ran the collision and propagation step for both lattices.
In the second variant, we tried to speed-up the calculations of nonlocal values by binding
the macroscopic fields to GPU textures and accessing them through these bindings. In the last
variant, we split the collision process into two separate kernels, one for every lattice.

The second variant yielded minimal speed-ups (on the order of a few percent) on old Tesla-generation
devices, which however did not carry over to Fermi and Kepler ones. We abandoned the idea as it introduced
unnecessary complexity to the code. The third approach using split kernels proved to be
the most efficient one. We were able to obtain 5.4\% (D2Q9, free energy) -- 53\% (D3Q19, free energy) speed-ups as compared to a monolithic
kernel, mainly due to the kernels using fewer registers and being able to achieve higher occupancy,
which hid the minor overhead introduced by a slightly increased number of global memory accesses.
As a side benefit, this approach also resulted in simpler code, so we decided to
standardize on it in the Sailfish framework. We also expect it to scale well to models requiring
more than 2 lattices.

\subsection{Distributed simulations}

In Sailfish, the mechanisms that support distributed simulations are very similar to those
supporting multiple subdomains on a single GPU or many GPUs on one computational node. Every subdomain
has a layer of \emph{ghost} nodes around it. These nodes do not participate in the simulation, but
are used for data storage for mass fractions leaving the subdomain or for macroscopic fields of
neighboring nodes located in remote subdomains. Once mass fractions are streamed into the ghost
nodes, we run additional CUDA kernels to collect the data leaving the subdomain into a linear
buffer in global memory, which is then transferred back to the host and sent to remote nodes
using a network link.

The subdomain is also split into two areas called
\emph{bulk} and \emph{boundary}, which are simulated via two separate kernel calls.
The boundary area is defined as all nodes belonging to CUDA thread blocks where at least one node
touches the subdomain boundary. The simulation step is first performed for nodes in the boundary area,
so that communication with remote subdomain runners can start as soon as possible and can overlap
in time with the simulation of the bulk of the subdomain. As a further optimization, we limit
the data sent between computational nodes exactly to the mass fractions that actually need to
be sent (e.g. if two nodes are connected along the X axis, then only the mass fractions corresponding
to discrete velocities with a positive X component will be transferred from the left subdomain to the
right one). We also make it possible to optionally compress the data before sending it to the
remote node, which can be helpful if the simulation is run on multiple machines with slow
network links between them, or when only a small fraction of nodes on the subdomain interface
plane is active.

\subsection{Impact of single precision}

Many GPUs are significantly faster when calculations are done in single precision as opposed
to the standard double precision which is typically used in scientific and engineering applications.
The speed-up factor can vary between 10 and 2, depending on the device model and generation.
The performance gap is smaller in newer devices (e.g. Kepler generation). Earlier works \cite{Kuznik20102380,Obrecht2011}
used the lid-driven cavity benchmark to verify that single precision calculations produce satisfactory
results.

Here, we use the 2D Taylor-Green decaying vortex flow -- a benchmark problem with known analytical solution
-- to study the accuracy of our LB implementation in both single (SP) and double (DP) precision. The simulations are
done using the LBGK relaxation model on a D2Q9 lattice. For single precision, both the standard formulation and the
round-off minimizing formulation \cite{DhumieresMrt2002} (SRO) are tested. The Taylor-Green
vortex can be described by the following system of equations:
\begin{align}
\begin{split}
    u_x(t) &= -u_0 \cos(x) \sin(y) e^{-2 \nu t} \\
    u_y(t) &=  u_0 \sin(x) \cos(y) e^{-2 \nu t} \\
    \rho(t) &= \frac{\rho}{4} \left(\cos(2x)  + \cos(2y) \right) e^{-4 \nu t}
\end{split}
\label{eq:taylor_green}
\end{align}
where $u_0$ is a velocity constant and $x, y \in [0;2\pi)$. \eqref{eq:taylor_green} can be easily
verified to satisfy the incompressible Navier-Stokes equations. We performed multiple runs of the
simulation on a $256^2$ lattice with periodic boundary conditions in both directions,
varying the viscosity $\nu$ but keeping the Reynolds number constant at $\Rey = 1000$.
Each simulation was run until it reached a specific point in physical time,
corresponding to $10^6$ iterations at $u_0 = 0.0005$.
We use the L2 norm of the velocity and density field difference to characterize the deviation of the numerical solution from the
analytical one:
\begin{equation}
	\epsilon = \sum_{\mathrm{nodes}}\sqrt{\frac{(u_x - \hat{u}_x)^2 + (u_y - \hat{u}_y)^2}{u_x^2 + u_y^2}}
	\label{eq:velocity_dev}
\end{equation}
where $\hat{u}_x$, $\hat{u}_y$ is the numerical solution.

\begin{figure}
	\centering
	\includegraphics[width=\textwidth]{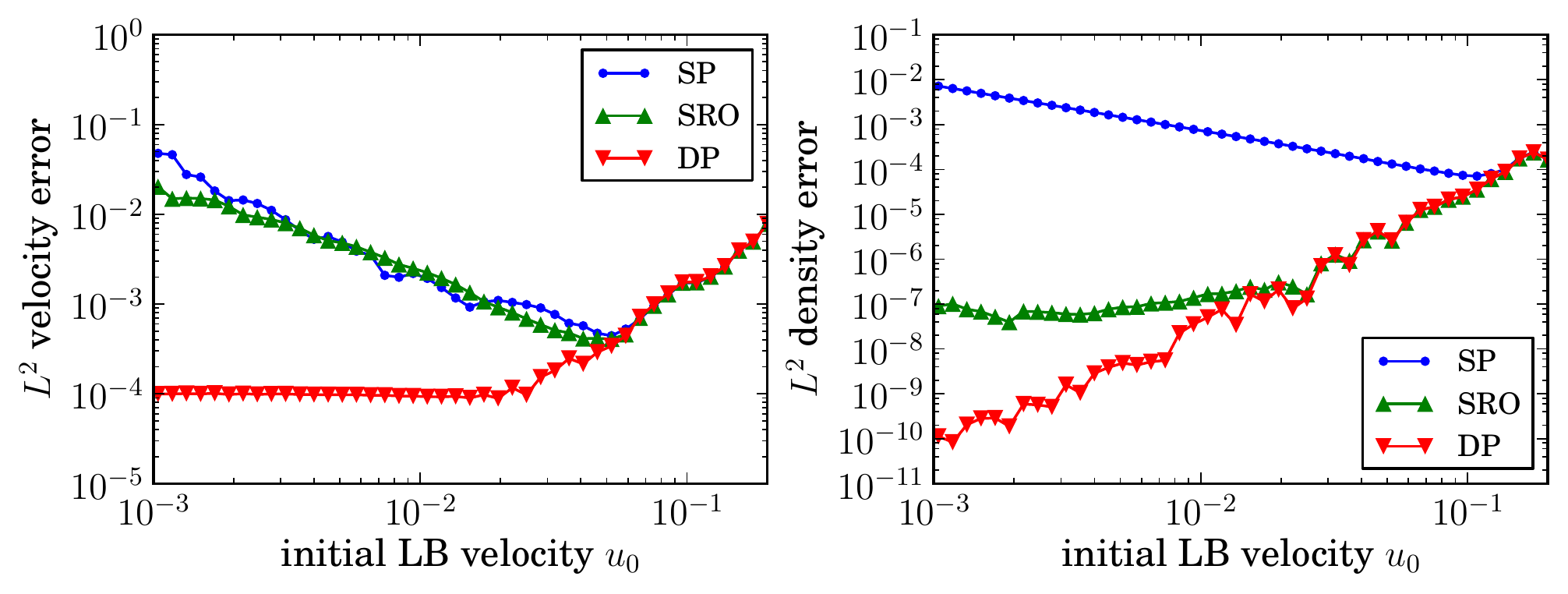}
	\caption{Solution error for the Taylor-Green vortex test case with the LBGK model on a D2Q9 lattice,
in single precision (SP), single precision with round-off minimization (SRO) and double precision (DP). Left
panel: velocity error. Right panel: density error.}
	\label{fig:precision}
\end{figure}

The results presented in~\figref{precision} illustrate interesting differences between double precision
and single precision calculations. In double precision, the error stays constant until $u_0 \approx 0.02$ and
raises quadratically for higher values of velocity, as could be expected from the $O(\mathrm{Ma}^2)$
accuracy of the LBGK model. In single precision however, lower speeds lead to higher error values.
The interplay between model accuracy and numerical round-off leads to a sweet spot around $u_0 \approx 0.05$,
where the errors are minimized. For higher speeds, there is no difference between double and single precision.
This result shows that the conventional wisdom that lower speeds always lead to better precision is not true
when applied to single precision simulations. Since many practical simulations are run at velocities
$0.05$ and higher for reasons of efficiency, it also explains why in many cases no differences are
observed between single and double precision results. The density error shows that the round-off
minimization model generates significantly better results than the standard single precision
implementation, and as such should be preferred in cases where the accuracy in the density field
is important.

\section{Performance}

All performance figures in the this section are quoted in MLUPS (Millions of Lattice-site Updates per Second).
The tests were run using CUDA Toolkit 5.0, PyCUDA 2013.1.1 on 64-bit Linux systems.

\subsection{Comparison of different models}

\begin{figure}
	\centering
	\includegraphics[width=\textwidth]{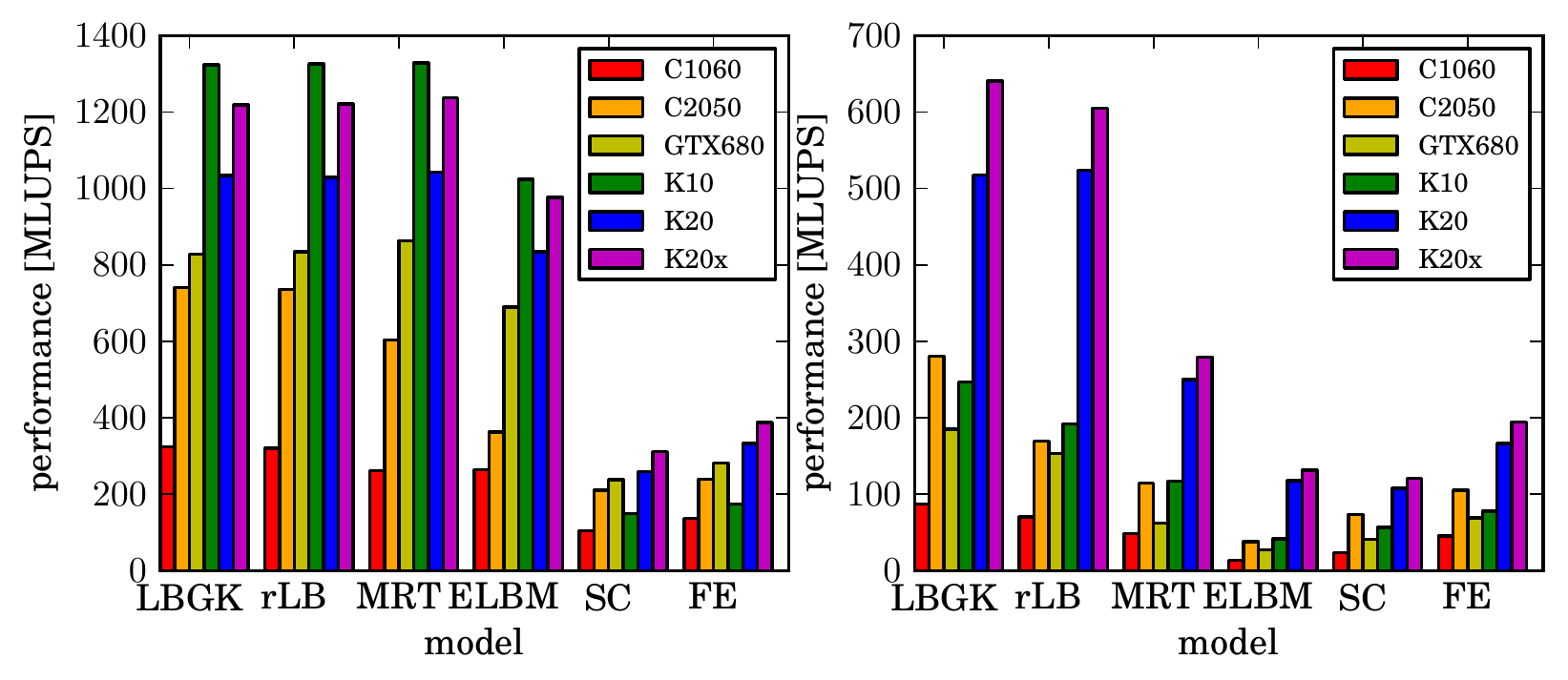}
	\caption{Performance comparison of different models using the D3Q19 lattice, AB memory access pattern.
Used acronyms: ELBM: entropic LBM, SC: Shan-Chen, FE: free energy.
All GPUs had ECC disabled. ELBM used intrinsic functions as described in section~\ref{sec:intrinsic}. Both logical GPUs were used for the K10 tests.
Left panel: single precision. Right panel: double precision.}
	\label{fig:perf_models}
\end{figure}

\begin{figure}
	\centering
	\caption{Performance of an LBGK simulation with the AA memory access pattern as a function of lattice type.
The lower performance for the K10 simulation for D3Q13 and D3Q15 lattices is caused by overhead in copying
the data between the two GPUs through the host (both logical GPUs were used for the K10 tests). Test case: Kida vortex (see text).
Left panel: single precision. Right panel: double precision.}
	\includegraphics[width=\textwidth]{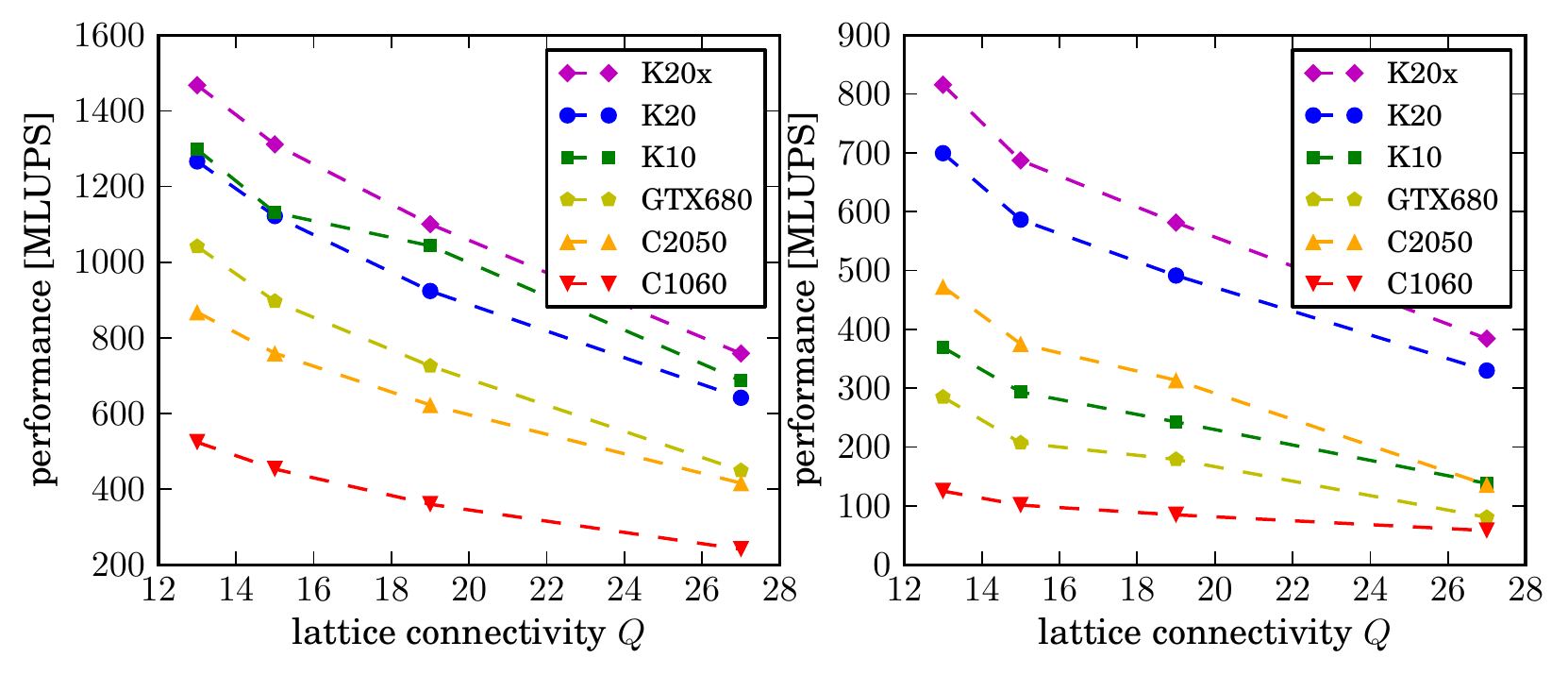}
	\label{fig:memory_grid}
\end{figure}

Single fluid models were benchmarked using a lid-driven cavity test case, with a $254^3$ lattice
at $\Rey = 1000$, using full bounce-back walls and equilibrium velocity boundary conditions for the lid.
To illustrate the impact of the entropy-stabilized time-stepping, the ELBM solver was also tested at $\Rey = 10000$.
The binary fluid models were benchmarked using a simple spinodal decomposition test case, where a uniform
mix of both fluid components fills the whole simulation domain ($254 \times 100 \times 100$) and periodic
boundary conditions are enabled in all directions. Whenever the domain size was too large to fit on a specific
GPU we tested, we reduced the size in the Z direction until it fit.

We find our results to be comparable or better to those reported by Habich et al.~\cite{Habich2012},
who used hardware very similar to ours (Tesla C2050 and C2070 differ only in memory size, according
to NVIDIA specifications). The performance of single precision simulations can be shown to be limited
by global memory bandwidth. We find that our code runs at $\sim80\%$ of the theoretical bandwidth as reported
in NVIDIA whitepapers, and close to 100\% of the real bandwidth measured by sample code from
the NVIDIA SDK. Double precision simulations run at $\sim60\%-80\%$ of the theoretical maximum. They are limited
by the double precision instruction throughput on the GPU on Fermi-class hardware and by memory bandwidth
in Kepler hardware. Overall, we find that the memory bandwidth
is a reliable indicator of expected performance for single precision simulations across all three
generations of NVIDIA devices (see also~\figref{memory_grid}, which shows that the simulation performance
is inversely proportional to the lattice connectivity $Q$ in single precision).


For double precision simulations on Fermi hardware, we have found increasing the L1 cache size to 48 kB,
disabling L1 cache for global memory accesses and replacing division operations by equivalent multiplication
operations to have a large positive impact on the performance of the code ($\sim 3.5$ speed-up in total).
The impact of the L1 cache is understandable if one considers the fact that
double precision codes use significantly more registers. This causes the compiler to spill some of them
over to local memory, accesses to which always go via the L1 cache. The larger the size of the unused part
of that cache, the more operations can execute without actually accessing the global memory.

This optimization strategy does not apply to Kepler-class devices, where both in single and double precision
we found that disabling the preference for L1 cache for the main computational kernel had a positive impact
on performance.

We also tested our code on lower end GPUs (mobile versions used in laptops) with compute capability 3.0,
where we found a significant speed up (up to 40\%) by using shuffle operations for in-warp propagation,
and limiting shared memory for data exchange between warps only. This optimization strategy does not work
with the higher end GPUs discussed in this paper, where the performance is limited by global
memory bandwidth already.

Overall, we unsurprisingly find that more recent GPUs perform noticeably better. The K10 is an interesting
option if only single precision is required as it delivers the highest overall performance at a level
higher than 1.3 GLUPS per board with D3Q19 and simple fluid models. For double precision, the K20x card
being the most advanced NVIDIA GPU available on the market when this paper is written, is a clear winner
performance-wise.

\subsection{Scaling on GPU clusters}

While the computational power provided by a single GPU is impressive, practical simulations often require
large domains, and for these the total size of GPU memory (a few GBs) is an important limitation.
The natural solution of this problem is to run a distributed simulation using multiple GPUs, which can
be physically located in a single computer or multiple computers in a network.

In order to measure performance of distributed simulations, we ran a 3D duct flow test case
(periodic boundary conditions in the streamwise Z direction, bounce-back walls at other
boundaries) on the Zeus cluster (part of the PLGRID infrastructure), consisting of
computational nodes with 8 M2090 GPUs and interconnected with an Infiniband QDR network.

\subsubsection{Weak scaling}

\begin{figure}
	\centering
	\includegraphics[width=\textwidth]{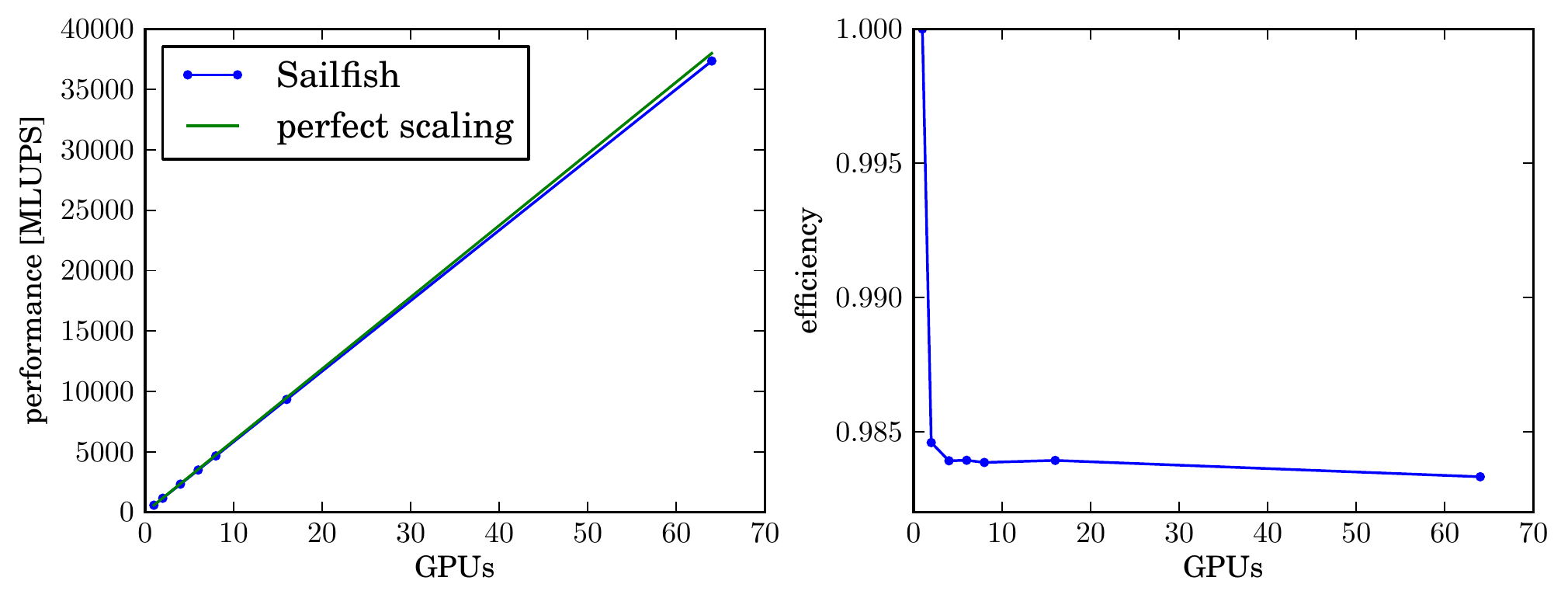}
	\caption{Weak scaling properties of the Sailfish code. The simulation was run on nodes with 8 M2090 GPUs, with one
subdomain used for every GPU. Test case: duct flow.
Left panel: absolute performance values. Right panel: efficiency fraction.}
	\label{fig:weak_scaling}
\end{figure}

The first test we performed measured weak scaling, i.e. code performance as a function
of increasing size of the computational domain. The domain size was
$254 \times 127 \times 512 \cdot N$, where $N$ is the number of GPUs. We used
the D3Q19 lattice, the AA memory access pattern, a CUDA block size of 128, and single
precision. \figref{weak_scaling}
shows excellent scaling up to 64 GPUs, which was the largest job size we were able to
run on the cluster. The 1.5\% efficiency loss takes place as soon as more than 1 subdomain
is used, and does not degrade noticeably as the domain size is increased. This small
efficiency loss could be further minimized by employing additional optimization techniques,
such as using peer-to-peer copies when the GPUs are located on the same host.

\subsection{Strong scaling}
\begin{figure}
	\centering
	\includegraphics[width=\textwidth]{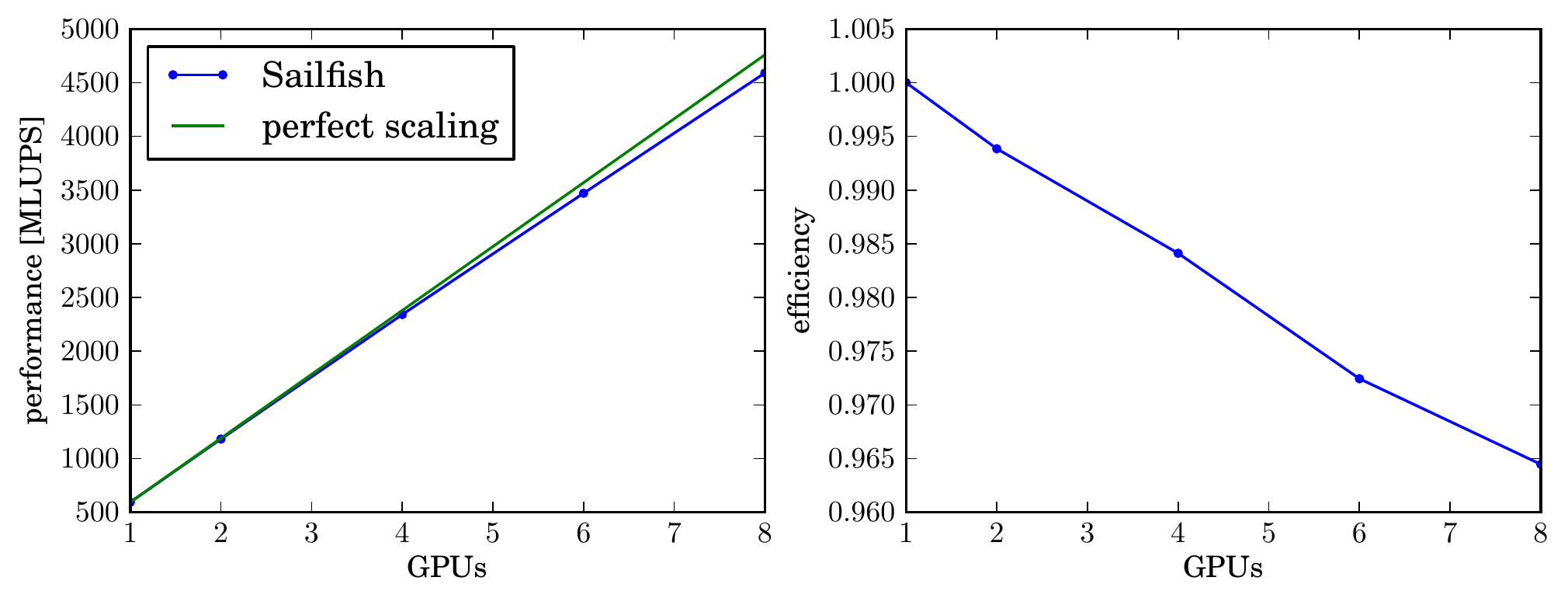}
	\caption{Strong scaling properties of the Sailfish code.
One subdomain was used for every M2090 GPU. Test case: duct flow.
Left panel: absolute performance values. Right panel: efficiency fraction.}
	\label{fig:strong_scaling}
\end{figure}

The second test we ran measured strong scaling, i.e. code performance with a constant
domain size, as a function of increasing number of GPUs. We used a $254 \times 127 \times 1664$ geometry
(largest domain size that fit within the memory of a single M2090 GPU) and other settings as in the weak
scaling test, and divided the domain into equal-length chunks along the Z axis as more GPUs
were used for the simulation. The results of this test are presented in \figref{strong_scaling}.
A slightly worse performance is visible in comparison to the weak scaling test, but
even when the simulation is expanded to run on 8 GPUs, only a 3.5\% efficiency loss can be
observed.

\begin{figure}
	\centering
	\includegraphics[width=0.5\textwidth]{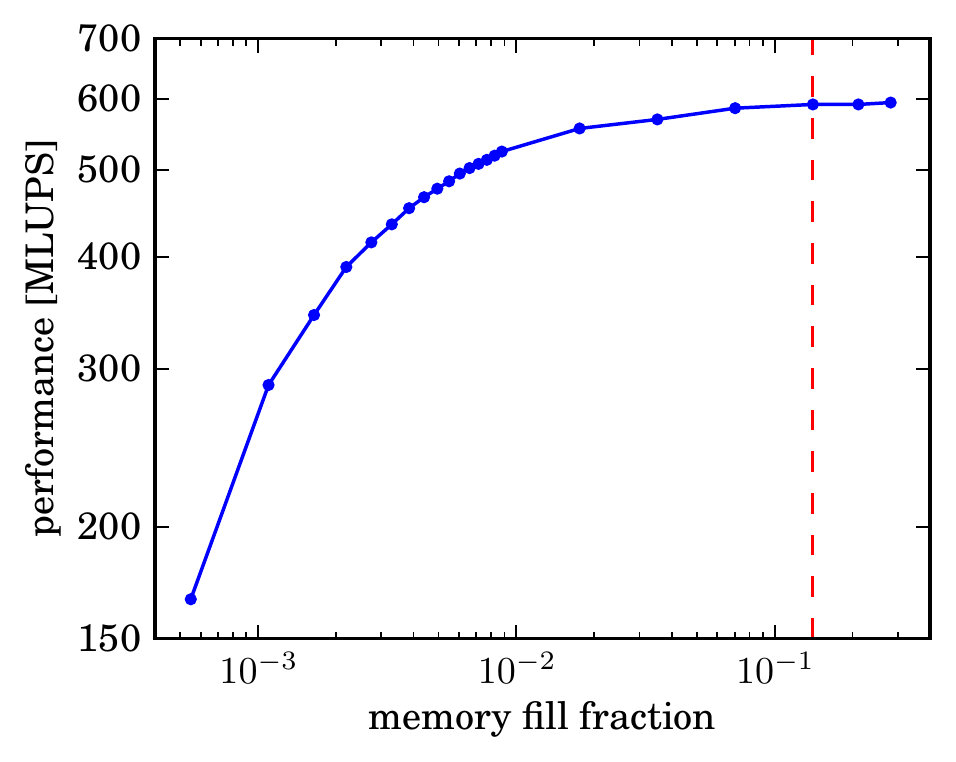}
	\caption{Duct flow simulation performance as a function of used memory fraction of a single M2090 GPU.
The vertical line shows the point at which the computational capabilities of the GPU are saturated. The geometry
was the same as in the strong scaling test and the memory fill fraction was controlled by varying the extent
of the domain in the Z direction. Very similar behavior is observed for K10 and K20 GPUs (not shown here).}
	\label{fig:min_domain_size}
\end{figure}

It should also be noted, that there is a minimum domain size below which performance quickly degrades
due to the overhead of starting kernels on the GPUs. This size can be seen in \figref{min_domain_size} to be 
about 14\% of the GPU memory or 8.2 M lattice nodes.

\subsection{Further optimization with intrinsic functions}
\label{sec:intrinsic}

CUDA GPUs provide an alternative hardware implementation of various transcendental functions
such as the exponent, logarithm or trigonometric functions. These functions, known
as intrinsic functions, are faster but less precise than their normal counterparts, especially
when their arguments do not fall within a narrow range specified for each function.

We analyzed the impact of these functions on the performance and precision of the LB models
that can take advantage of them, namely the Shan-Chen model with a non-linear pseudopotential
and the entropic LBM. With ELBM the use of intrinsic functions, together with the FMAD (fused
multiply-add) instruction yields a speed-up of $\sim43$\% without any noticeable impact on
the correctness of the results (in terms of global metrics such as the total kinetic energy, enstrophy
and the kinetic energy spectrum). While testing these optimizations, we also found that the FTZ
(denormalize to 0) option of the CUDA compiler causes the ELBM simulation to crash.

With the proposed optimizations, the performance of ELBM is at 72\% of the LBGK performance,
making it a very interesting alternative for some simulations.

For the Shan-Chen model with a nonlinear pseudopotential we saw 17-20\% speed-ups in 2D and 3D,
with relative changes in the density fields smaller than 1.5\% after 10000 steps.

Unfortunately, the same approach does not yield speed-ups in double precision, as most 
intrinsic functions are available in single precision only.
\section{Validation}

In order to validate our implementation of the LBM, we performed simulations for
four classical computational fluid dynamics test cases and compared our results
with those published in the literature.

\subsection{Lid-driven cavity}

The lid-driven cavity geometry consists of a cube cavity with a face length $L$. The geometric center
of the cavity is located at the origin of the coordinate system. The cube face at $x = - L/2$ moves
tangentially in the y-direction with a constant velocity $v$, while all other faces are no-slip walls.
We carried out simulations of this problem at $\Rey = 1000$ with various LB models (BGK, MRT, regularized BGK, ELBM)
using a $201^3$ D3Q19 lattice, full-way bounce-back for no-slip walls, and the regularized velocity
boundary condition with $u_y = 0.05$ for the moving wall.

\begin{figure}
	\centering
	\includegraphics[width=0.5\textwidth]{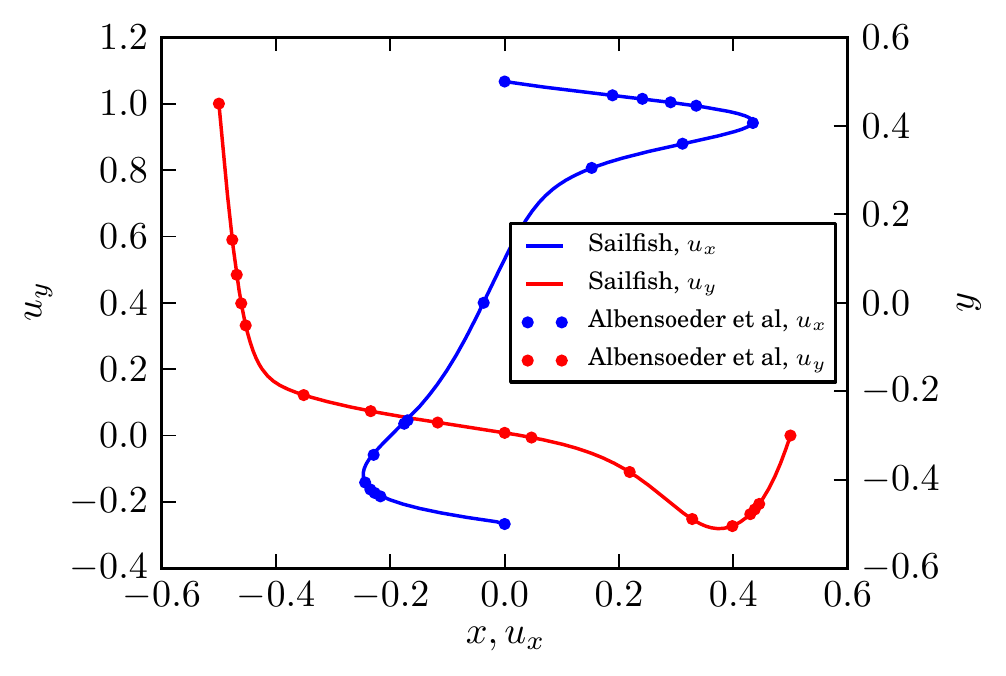}
	\caption{Lid-driven cavity velocity profiles $u_y(x, 0, 0)$ and $u_x(0, y, 0)$. Round dots: data from Tables~5 and~6 in Albensoeder et al.~\cite{Albensoeder2005536}
		Solid line: results from Sailfish simulations on a $201^3$ lattice, after $2 \cdot 10^5$ steps using LBGK, MRT, regularized BGK and ELBM in single and double precision. The results from
		all Sailfish simulations are in agreement to within the width of the line on the plot. LB results are rescaled using: $\vec{u} = \vec{u}_\mathrm{LB} / 0.05$.}
	\label{fig:ldc-comparison}
\end{figure}

Our results (both in single and double precision) agree with those published by Albensoeder et al.~\cite{Albensoeder2005536} (see
\figref{ldc-comparison})

\subsection{Kida vortex}

The Kida vortex flow is a free decay from the initial conditions~\cite{Kida1985}:
\begin{align*}
	u_x(\vec{x}, 0) &= u_0 \sin x \left( \cos 3 y \cos z - \cos y \cos 3 z \right) \\
	u_y(\vec{x}, 0) &= u_0 \sin y \left( \cos 3 z \cos x - \cos z \cos 3 x \right) \\
	u_z(\vec{x}, 0) &= u_0 \sin z \left( \cos 3 x \cos y - \cos x \cos 3 y \right)
\end{align*}
defined on a cubic domain with face length $2 \pi$, with periodic boundary conditions in all directions.
To validate our code, we performed simulations for $\Rey = N u_0 / \nu = 4000$, $1.28 \cdot 10^4$, and $1.28 \cdot 10^5$ and
compared them with results published in \cite{ChikatamarlaKidaDNS} and \cite{PhysRevE.75.036712}, respectively.
The simulations were run using $u_0 = 0.05$ on a $700^3$ grid ($350^3$ for $\Rey = 4000$), using both single and double precision (with
no noticeable difference between them). The $\Rey = 4000$ and $\Rey = 1.28 \cdot 10^4$ cases were investigated
using the LBGK, MRT, regularized LBGK, Smagorinsky-LES and entropic models. At $\Rey = 1.28 \cdot 10^5$, only simulations using
the entropic model and the Smagorinsky subgrid model (with $C_S = 0.1$) remained stable. During the simulation time kinetic
energy $E = \frac{1}{2 V} \int d^3 x \vec{v}^2$ and enstrophy $\Omega = \frac{1}{2 V} \int d^3 x (\vec{\nabla} \times \vec{v})^2$,
where $V$ is the volume of the simulation domain, were tracked directly on the GPU for optimal efficiency. Vorticity was computed using the 
first order central difference scheme in the bulk of the fluid and forward/backward differences at subdomain boundaries.
\begin{figure}
	\centering
	\includegraphics[width=\textwidth]{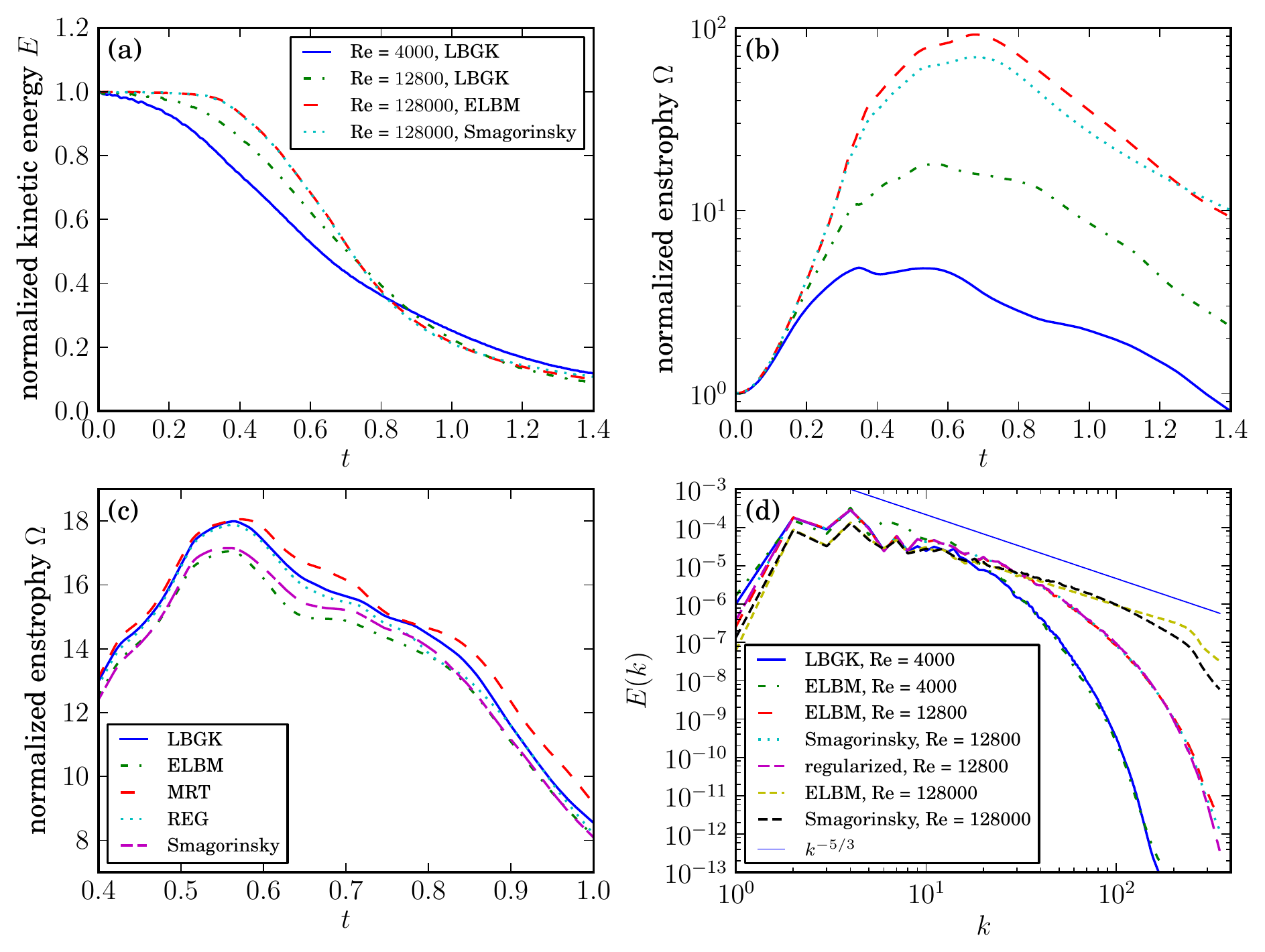}
	\caption{(a) evolution of normalized kinetic energy, (b) evolution of normalized enstrophy, (c) evolution of normalized enstrophy at $\Rey = 12800$ for various collision models,
(d) kinetic energy spectrum for selected collision models and Reynolds numbers. For panels (a)-(c) time is rescaled assuming a domain size of $(2 \pi)^3$ and $u_0 = 1$.}
	\label{fig:ke-ens}
\end{figure}

For $\Rey=4000$ all four models gave the same results (\figref{ke-ens}). At $\Rey = 1.28 \cdot 10^4$ some minor differences
are visible, particularly in the evolution of enstrophy. Its peak value is slightly underpredicted by both models that
locally modify effective viscosity (Smagorinsky, ELBM) (see \figref{ke-ens}(c)).  At $\Rey = 1.28 \cdot 10^5$, the differences are more pronounced
and we observe that the Smagorinsky model underpredicts the absolute value of peak enstrophy. The kinetic energy spectrum shown on \figref{ke-ens}(d)
was computed as $E(k) = \sum_{k \leq k' < k+1} \hat{u}(k)^2$, for $k = 0, 1, 2, \ldots$. A good agreement is visible in comparison to
the Kolmogorov scaling $k^{-5/3}$, especially for the high Reynolds number cases. All collision models lead to similar spectra, with
ELBM at $\Rey = 4000$ predicting a slightly higher value around $k = 10$ than LBGK or other models, and with ELBM keeping a slightly
flatter spectrum for high $k$ values at $\Rey = 1.28 \cdot 10^5$.  In all cases the simulation results show the same features as those
discussed in previous papers on this topic~\cite{Kida1985,ChikatamarlaKidaDNS,PhysRevE.75.036712}.

\subsection{Binary Poiseuille flow}

\begin{figure}
	\centering
	\includegraphics[width=\textwidth]{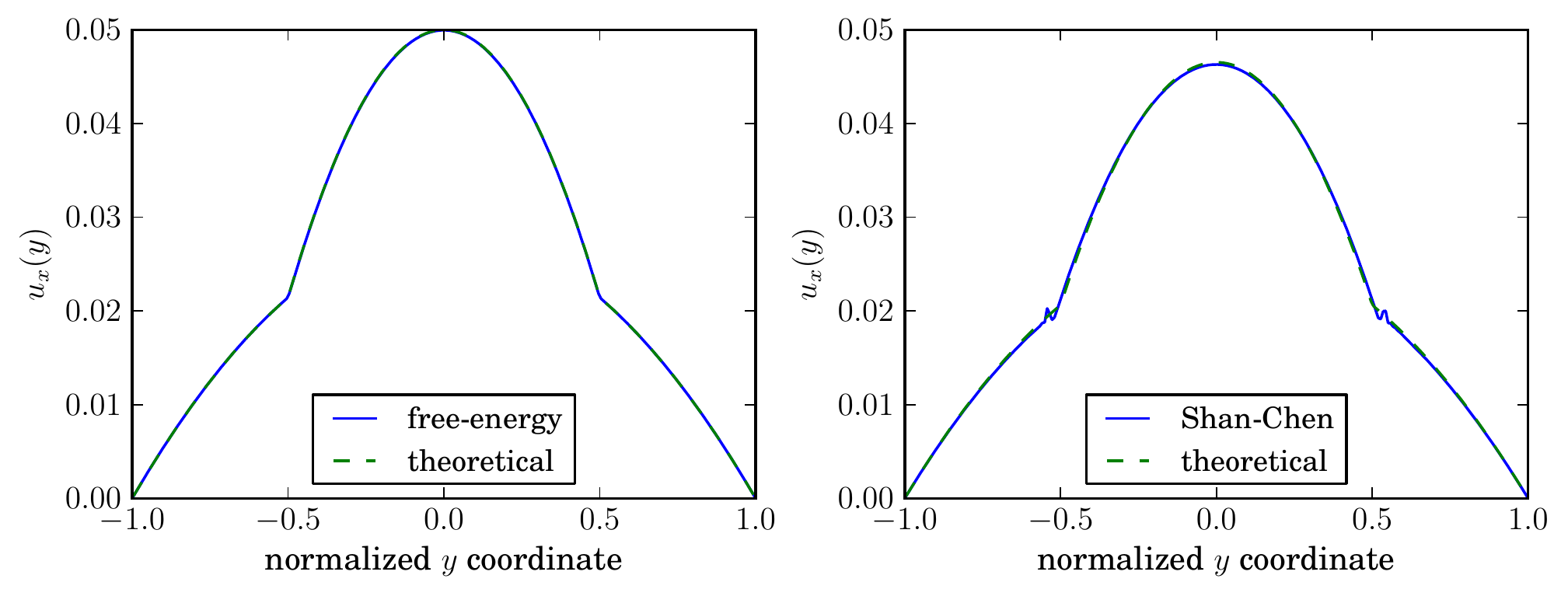}
	\caption{Comparison between simulated and theoretical velocity profiles for the binary Poiseuille flow.
Left panel: Results for the free energy model.
Right panel: Shan-Chen results. The effective viscosity ratio was $3.7985$ due to mixing between the two fluid components. The free energy
model shows better agreement with the theoretical profile due to a thinner interface.}
	\label{fig:poiseuille_comparison}
\end{figure}

To verify the binary fluid models we consider a 2D Poiseuille flow in the $x$-direction. No-slip walls
are imposed at $y = \pm L$ using the half-way bounce-back boundary conditions, and periodic boundary conditions
are used in the $x$-direction. A body force $G$ drives the flow. In the core flow region ($|y| \leq L / 2$)
a fluid of viscosity $\nu_1$ is placed, while in the boundary flow region ($L / 2 < |y| < L$) the viscosity
of the fluid is $\nu_0$. The analytical solution for this case can be expressed as:
\begin{equation}
u_x(y) = \begin{cases}
	\frac{G}{2 \rho \nu_0} \left(L^2 - y^2\right) &\mbox{if }  L / 2 < |y| < L \\
	\frac{G L^2}{8} \left( \frac{3}{\rho \nu_0} + \frac{1}{\rho \nu_1} \left( 1 - 4 \frac{y^2}{L^2} \right) \right)  &\mbox{if } |y| \leq L/2.
\end{cases}
	\label{eq:poiseuille_solution}
\end{equation}
We run the simulation on a $64 \times 256$ grid with $\nu_0 = 1/6$, and $\nu_1 = 1/24$. The simulation starts with
 $u_x = 0$ in the whole domain and we let the flow reach the stationary state on its own. The free energy simulation
was run with $\Gamma = 25$, $\kappa = 10^{-4}$, and $A = 32 \cdot 10^{-4}$, while for the Shan-Chen model, $G = 1.5$
was used. The parameters were chosen to ensure that the interface remains as sharp as possible without destabilizing
the simulation. The Exact Difference Method was used to introduce body forces. The Shan-Chen model permits residual
mixing between the fluid components, so the effective density and viscosity were calculated as $\rho = \rho_0 + \rho_1$ and
$\nu \rho = \rho_0 \nu_0 + \rho_1 \nu_1$, respectively. \figref{poiseuille_comparison}
illustrates the good agreement of the simulation results with \eqref{eq:poiseuille_solution}.

\subsection{Capillary waves}

In order to verify the binary fluid models also in a dynamic case, we simulate the decay of the amplitude
of a capillary wave. The boundary conditions of the system are the same as in the
binary Poiseuille flow case, but we now use a larger $512 \times 512$ lattice in order to accommodate higher
frequency waves. The region $y > 0$ is filled with one component (A) and the region
$y < 0$ is filled with another component (B). For simplicity, we choose both components to have the
same viscosities $\nu = 1/18$. The interface between the two fluids is initialized
to a sinusoidal curve of wavelength $\lambda$ chosen such that an integer number of wavelengths
fits exactly in the simulation domain. The interface is then allowed to relax naturally with no
external forces, resulting in a damped capillary wave. At each timestep of the simulation, the
height of the interface at $x = L - \lambda / 4$ is recorded. In order to recover the frequency
of the wave, an exponentially damped sine function is fit to the interface height data. For the
Shan-Chen model, we used $G = 0.9$ and for the free energy model we used $A = 0.02$, $\kappa = 0.04$,
and $\Gamma = 0.8$.
\begin{figure}
	\centering
	\includegraphics[width=0.5\textwidth]{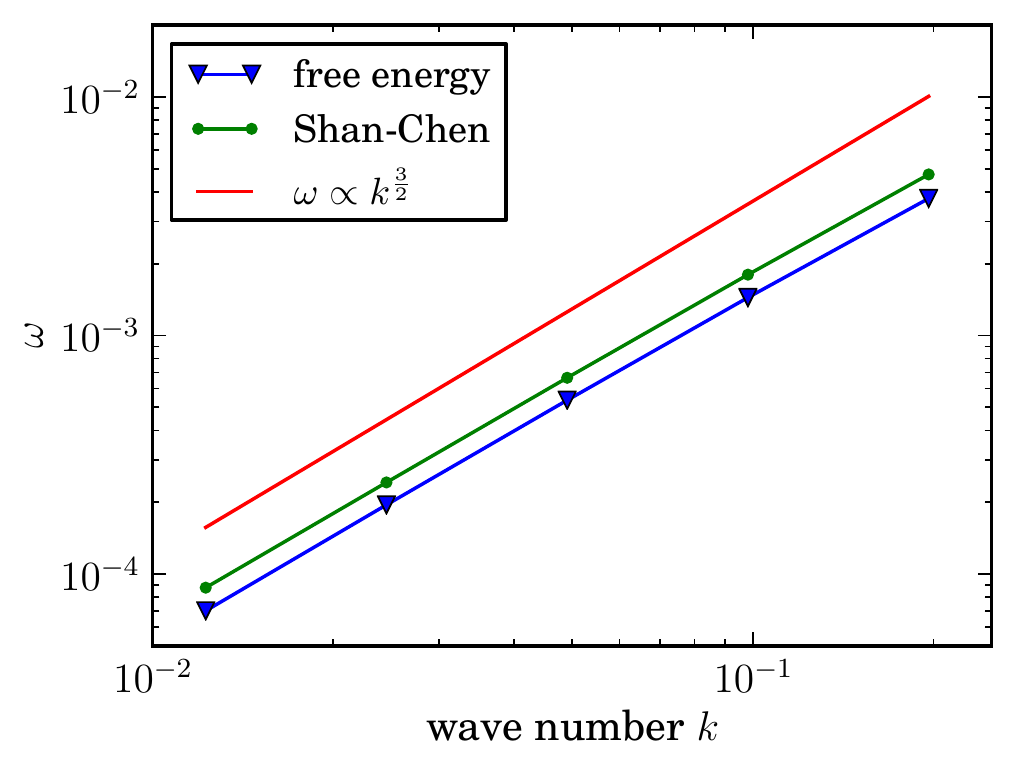}
	\caption{Theoretical and measured dispersion relation $\omega(k)$ for the capillary wave
	for the free energy and Shan-Chen models.}
	\label{fig:capillary}
\end{figure}
As expected \cite{Langaas_Yeomans,chin_binary_shan_chen}, the dispersion relation shows a power law
form $\omega \propto k^{3/2}$ for both the Shan-Chen and free energy models (see \figref{capillary}).

\section{Conclusions}

In the previous sections we have demonstrated our Sailfish code as a flexible
framework for implementing lattice Boltzmann models on modern graphics processing units. With
novel optimization techniques for complex models it provides a very efficient
tool for a wide range of simulations. We hope that our observations collected while running
the presented benchmarks will serve as a guideline in the choice of both LB models and
computational hardware for users of Sailfish and of other similar codes.

For single precision simulations, we advocate a careful choice of parameters and
correctness testing via comparisons to similar test cases in double precision.
While all of our benchmark problems did not show any noticeable differences
between single and double precision, the Taylor-Green test case clearly demonstrates
that these do exist and can significantly impact the results if the simulation
is in the slow velocity regime. Whenever possible, the round-off minimizing model should be
used to reduce precision losses without any impact on performance.

While the capabilities of Sailfish are already quite extensive, much work remains to be done.
Among the most important remaining tasks, we mention ongoing efforts to implement
a hierarchical lattice, allowing for local grid refinement and a fluid-structure interaction
model based on the immersed boundary method. Since the code is freely available under
an open source license, we would like to invite the reader to participate in its
development and contribute new enhancements according to their interests.

\subsection{Acknowledgments}

This work was supported by the TWING project co-financed by the European Social Fund, as well as
in part by PL-Grid Infrastructure. M.J. thanks S.~Chikatamarla and I.~Karlin for useful discussions and sharing
reference data from their simulations. The authors would also like to thank NVIDIA for providing
hardware resources for development and benchmarking.

\bibliographystyle{elsarticle-num}

\end{document}